\documentclass[journal,twocolumn,10pt]{IEEEtran}
\normalsize
\usepackage[noadjust]{cite}
\usepackage{filecontents}
\usepackage{color,colortbl}
\usepackage[pdftex]{graphicx}
\usepackage{verbatim} 
\usepackage{epstopdf}
\usepackage[cmex10]{amsmath}
\usepackage[tight,footnotesize]{subfigure}
\usepackage{pifont}
\usepackage{algorithm}
\usepackage{algpascal}
\usepackage{algpseudocode}
\usepackage{lipsum}
\usepackage{amssymb}
\definecolor{Gray}{gray}{0.9}
\newcommand{\E}{{\rm I\kern-.3em E}}

\begin{document}
\title{Low-Complexity Sub-band Digital Predistortion for Spurious Emission Suppression in Noncontiguous Spectrum Access}
\author{Mahmoud~Abdelaziz,~\IEEEmembership{Student~Member,~IEEE,}
        Lauri~Anttila,~\IEEEmembership{Member,~IEEE,} 
				Chance~Tarver,~\IEEEmembership{Student~Member,~IEEE,}
				Kaipeng~Li,~\IEEEmembership{Student~Member,~IEEE,}
				Joseph~R.~Cavallaro,~\IEEEmembership{Fellow,~IEEE,}
        Mikko~Valkama,~\IEEEmembership{Senior~Member,~IEEE}        
\thanks{Mahmoud~Abdelaziz, Lauri~Anttila, and Mikko~Valkama are with the Department of Electronics and Communications Engineering, Tampere University of Technology, Tampere, Finland. 

Chance~Tarver, Kaipeng~Li, and Joseph~R.~Cavallaro are with the Department of Electrical and Computer Engineering, Rice University, Houston, TX.} 
%\thanks{This work was supported by the Finnish Funding Agency for Technology and Innovation (Tekes) under the projects "Cross-Layer Modeling and Design of Energy-Aware Cognitive Radio Networks (CREAM)", and "Future Small-Cell Networks using Reconfigurable Antennas (FUNERA)". The work was also funded by the Academy of Finland under the projects 251138 "Digitally-Enhanced RF for Cognitive Radio Devices" and 284694 "Fundamentals of Ultra Dense 5G Networks with Application to Machine Type Communication", and by the Linz Center of Mechatronics (LCM) in the framework of the Austrian COMET-K2 programme.}%
\thanks{This work was supported by the Finnish Funding Agency for Technology and Innovation (Tekes) under the project ``Future Small-Cell Networks using Reconfigurable Antennas (FUNERA)", and by the Linz Center of Mechatronics (LCM) in the framework of the Austrian COMET-K2 programme. The work was also funded by the Academy of Finland under the projects 288670 ``Massive MIMO: Advanced Antennas, Systems and Signal Processing at mm-Waves", 284694 ``Fundamentals of Ultra Dense 5G Networks with Application to Machine Type Communication", and 301820 ``Competitive Funding to Strengthen University Research Profiles". This work was also supported in part by the US National Science Foundation under grants ECCS-1408370, CNS-1265332, and ECCS-1232274.}%
}
\maketitle
\begin{abstract}
%Noncontiguous transmission is one of the key features of recent wireless communication systems due to spectrum scarcity and requirements for higher data rates. However, the flexibility in such transmissions puts a lot of design challenges on the transmitter side, especially for mobile devices and small-cell base-stations. Additionally, multicarrier transmissions used in most recent wireless communication systems exhibit very large peak-to-average power ratio (PAPR). High PAPR noncontiguous transmissions combined with high power efficiency requirements in such small devices, pose a real challenge on the digital front end design due to the transmitter nonlinearity. Severe spectral emissions can be generated, which occur in adjacent channels and can potentially interfere with existing neighboring transmissions, thereby deteriorating the reliability of the network. Digital predistortion (DPD) is one of the most effective solutions for mitigating transmitter nonlinearity. 
Noncontiguous transmission schemes combined with high power-efficiency requirements pose big challenges for radio transmitter and power amplifier (PA) design and implementation. Due to the nonlinear nature of the PA, severe unwanted emissions can occur, which can potentially interfere with neighboring channel signals or even desensitize the own receiver in frequency division duplexing (FDD) transceivers. In this article, to suppress such unwanted emissions, a low-complexity sub-band DPD solution, specifically tailored for spectrally noncontiguous transmission schemes in low-cost devices, is proposed. The proposed technique aims at mitigating only the selected spurious intermodulation distortion components at the PA output, hence allowing for substantially reduced processing complexity compared to classical linearization solutions. Furthermore, novel decorrelation based parameter learning solutions are also proposed and formulated, which offer reduced computing complexity in parameter estimation as well as the ability to track time-varying features adaptively. Comprehensive simulation and RF measurement results are provided, using a commercial LTE-Advanced mobile PA, to evaluate and validate the effectiveness of the proposed solution in real world scenarios. The obtained results demonstrate that highly efficient spurious component suppression can be obtained using the proposed solutions.

% , is presented and compared to existing solutions. The proposed technique can be used to suppress arbitrary spurious intermodulation distortions adaptively according to the imposed limitations on the spectral emissions with little or no effect on the main carriers. Thus, it can be considered a step towards cognitive DPD where the hardware resources and computational complexity of the DPD are optimized according to the needs of the device. This is particularly beneficial for mobile devices and small-cell base-stations where hardware and power resources are limited compared to macro basestations.  Practical measurements using real RF equipments and a commercial mobile PA are also presented to further validate the effectiveness of the proposed solution in real world scenarios. 
\end{abstract}


\section{Introduction}
\label{sec:introduction}
Spectrum scarcity and ever-increasing data rate requirements are the two main motivating factors behind introducing carrier aggregation (CA) in modern wireless communication systems, such as 3GPP LTE-Advanced \cite{LTE_A_Parkvall,3GPP}. In CA transmission, multiple component carriers (CCs) at different RF frequencies are adopted simultaneously, either within the same RF band (intraband CA), or at different RF bands (interband CA) \cite{TransmitterArchitectureCA}. While interband CA naturally leads to a \emph{non-contiguous transmit spectrum}, this can also happen in the intraband CA case when the adopted component carriers are located in a non-contiguous manner. Adopting non-contiguous CA provides lots of flexibility in RF spectrum use, but also poses substantial challenges in the transmitter design, especially in lower-cost devices such as LTE-Advanced user equipment (UE).
Moreover, multicarrier type modulations with very large peak-to-average power ratio (PAPR) are used in most state-of-the-art wireless communication systems. When combined with noncontiguous transmission schemes and high power-efficiency requirements, controlling the transmitter unwanted emissions due to power amplifier (PA) nonlinearity becomes a true challenge \cite{CommagCA_LTE,3GPP_CA_Emissions_1,3GPP_CA_Emissions_2}. 

In \cite{TransmitterArchitectureCA}, a comparison of different transmitter architectures for CA based transmission schemes was presented, showing that it is more power efficient to combine the CCs before the PA, and thus use only a single PA, instead of having a separate PA for each CC. This is also technically feasible, especially in the intraband CA case. The power savings can be quite significant since the overall transmitter efficiency is dominated by the PA \cite{TransmitterArchitectureCA}, and is particularly critical in mobile terminals and other lower-cost devices, such as small-cell base-stations, with limited  computing and cooling capabilities \cite{TI_DPD,GreenComm}. However, the power efficiency gained from using a single PA for all CCs comes at the expense of more severe unwanted emissions, stemming from the PA nonlinear characteristics, that can occur at adjacent channels or bands, or even at own receiver band in FDD transceiver case \cite{CommagCA_LTE,3GPP_CA_Emissions_1,3GPP_CA_Emissions_2,3GPP_duplexer}.
%Furthermore, if the CC combiner after the PA had, e.g., a 3dB loss, $50\%$ of the PA output power would be wasted. The power efficiency considerations are even more
%Moreover, a major design principle in small-cell base-stations and mobile devices is to reduce the cost of the final product in terms of the bill of materials as well as the running cost. Consequently, low cost and small size RF components are highly desirable in such devices \cite{TI_DPD}, leading to even more unwanted emissions.

%Such unwanted emissions can potentially interfere with other existing transmissions, thereby deteriorating the reliability of the network \cite{TransmitterArchitectureCA,CommagCA_LTE}. 

The levels of different unwanted emissions are, in general, regulated by the standardization bodies and ITU-R \cite{ITU}. It has been recently demonstrated that in the context of LTE-Advanced mobile transmitter with noncontiguous CA or multicluster type transmission, the PA nonlinearities lead to spurious emissions that can seriously violate the given spectrum and spurious emission limits if not properly controlled \cite{3GPP,CommagCA_LTE,3GPP_CA_Emissions_1,3GPP_CA_Emissions_2}. Furthermore, in FDD devices, in addition to violating the general spurious emission limits, the generated spurious components can also overlap with the device's own receive band, causing own receiver desensitization \cite{CommagCA_LTE,3GPP_duplexer,RX_Desens_Cancellation_Chao,RX_Desens_Cancellation_Adnan}. One obvious solution to decrease the levels of unwanted emissions is to back off the transmit power from its saturation region. In 3GPP UE terminology, this is known as Maximum Power Reduction (MPR), and MPR values up to 16 dB are allowed in some use cases for mobile terminals \cite{3GPP}. However, this approach will end up yielding a significantly lower PA efficiency as well as a substantial reduction in the network coverage. There is thus a clear need for alternative PA linearization solutions which do not have such drastic adverse effects.

%Digital predistortion (DPD) is one of the most effective solutions for mitigating transmitter nonlinear distortion. However, most of the current DPD literature is focused on adopting DPD in macro base-stations or other high-end infrastructure nodes, where the available computing resources are high. Only few works have so far considered DPD in mobile terminals, or other lower-cost devices, where the computing power is much more limited. Consequently, our main focus in this paper is on developing low cost and highly efficient DPD solutions suitable for lower-cost devices with limited computing capabilities, with special focus on non-contiguous transmission scenarios.

Digital predistortion (DPD) is, in general, one of the most effective solutions for mitigating transmitter nonlinear distortion. There have been substantial research efforts in the past 20 years in developing ever more efficient and elaborate DPD techniques, mostly for high-end macro base-station type of devices. The conventional DPD approaches seek to linearize the full composite transmit signal, and we thus refer to such solutions as full-band DPD in this article. There have also been a handful of recent works on efficient concurrent linearization techniques for multi-band transmitters that employ only a single PA \cite{S.A.BassamOct.2011,RoblinNov2013,SingleFB_DPD}. These works assume that the component carriers are separated by a large distance such that the spurious emissions are filtered out by the transmit RF filter, and hence linearization of only the main carriers is pursued. 

\textcolor{black}{Complementary to such methods, the scope of this paper is to introduce a low-complexity DPD solution for suppressing the spurious emissions in non-contiguous transmission cases, while not concentrating specifically on the linearization of the main component carriers. Hereafter, we will refer to such linearization solutions as \textit{sub-band DPD}. This approach is motivated by the following two factors.}
First, the emission limits in the spurious region are generally stricter than in the spectral regrowth region around the component carriers, and are thus more easily violated. This has been recognized also in 3GPP recently, in the context of intraband noncontiguous carrier aggregation \cite{3GPP,CommagCA_LTE,3GPP_CA_Emissions_1,3GPP_CA_Emissions_2}. Moreover, even in interband carrier aggregation scenarios in FDD devices, some of the spurious components can be hitting the own RX band causing own receiver desensitization \cite{RX_Desens_Cancellation_Chao}. 
Second, by concentrating the linearization efforts to the most critical spurious emissions only, the processing and instrumentation requirements can be significantly relaxed, thus potentially facilitating the DPD processing also in mobile terminals and other lower-cost devices. This applies not only to the DPD main path processing complexity but also to the feedback receiver instrumentation complexity which can also be substantially reduced. 

%Thus, the idea is to develop a reduced complexity, real-time adaptive DPD that can significantly enhance the power efficiency and prolong the battery life of these small devices.
There have been some recent studies in the literature that consider the mitigation of the spurious emissions explicitly. In \cite{P.RoblinJan.2008,J.KimJan.2013,S.A.BassamAug.2012}, such processing was added to complement a concurrent linearization system while assuming a frequency-flat PA response within the spurious sub-band. In \cite{P.RoblinJan.2008,J.KimJan.2013}, the DPD parameter estimation was done offline, based on extracting the quasi-memoryless PA parameters using large-signal network analyzer (LSNA) measurements, and covering up to fifth-order processing. In \cite{S.A.BassamAug.2012}, a memoryless least-squares fit between the observed intermodulation distortion (IMD) at the considered sub-band and certain basis functions was performed. The basis functions were generated using a wideband composite carrier baseband equivalent signal, followed by sub-band filtering, implying a very high sample rate and processing complexity, especially for widely spaced carriers.
% However, a low sample rate observation receiver was used to extract the IMD at the desired sub-band to be used for DPD parameter estimation.
The estimated and regenerated IMD was then applied at the PA input, oppositely phased, such that it is canceled at the PA output. 
\textcolor{black}{
In \cite{ICASSP2014,CROWNCOM2014}, on the other hand, third-order nonlinearities at the IM3 sub-bands were specifically targeted, through explicit, low-rate, behavioral modeling of the baseband equivalent IM3 sub-band emissions. Furthermore, the parameter estimation of the DPD was based on a closed-loop feedback with a decorrelation-based learning rule, which was shown to have better linearization performance compared to the third-order inverse solution in \cite{P.RoblinJan.2008} in terms of the spurious emission suppression. Moreover, an FPGA implementation of this third-order decorrelation-based sub-band DPD was presented in \cite{ASILOMAR_ABDELAZIZ}, demonstrating fast and reliable performance under real time constraints. 
% All the above works assumed a special DPD structure for the spurious IMD emissions, which we here refer to as sub-band DPD. 
A recent overview article \cite{COMMAG_ABDELAZIZ} highlighted the main principles and advantages of such low complexity sub-band DPD solutions, while not concentrating on the details of the DPD processing or parameter estimation and adaptation algorithms at technical level. 
%The main objective in \cite{COMMAG_ABDELAZIZ} was to spark discussion, raise awareness, and catalyze further research in the interesting area of reduced complexity sub-band DPD solutions for non-contiguous transmissions, in particular for mobile devices. 
In \cite{TCOM_ABDELAZIZ}, on the other hand, a flexible full-band DPD solution was also proposed by the authors which can optimize the DPD coefficients to minimize the nonlinear distortion at a particular frequency or sub-band in the out-of-band or spurious regions. However, like other full-band DPD techniques, it requires very high sampling rates in the transmitter and feedback receiver when the carrier spacing between the CCs increases.}

In this article, we extend the elementary third-order IM3 sub-band DPD solution, proposed by the authors in \cite{ICASSP2014,CROWNCOM2014}, in two ways. First, the third-order IM3 sub-band DPD is extended to incorporate higher-order processing based on explicit modeling of the higher-order spurious components at the IM3 sub-band. This will enhance the IM3 spurious emission suppression considerably. Furthermore, we also extend the sub-band DPD solution to include higher-order sub-bands, i.e., IM5, IM7, etc., thus offering more flexibility and linearization capabilities beyond the basic IM3 sub-band. All the proposed solutions are derived for wideband nonlinear PAs with memory. Furthermore, we also formulate novel decorrelation-based parameter estimation methods, covering both sample-adaptive and block-adaptive learning rules, to efficiently identify the needed DPD parameters with low complexity. The proposed learning solutions are also shown to offer better performance than the earlier proposed third-order or fifth-order inverse based methods. We also provide comprehensive simulation and RF measurement results, using a commercial LTE-Advanced mobile PA, to evaluate and validate the effectiveness of the proposed solutions in real world scenarios.

%Moreover, the PA memory effects are considered in the sub-band DPD learning using a simple multi-tap adaptive filter structure. A block-adaptive version of the decorrelation-based sub-band DPD is also introduced in this paper. The block-adaptive solution aims at facilitating the implementation of the proposed solution in real-time environments with arbitrary loop delays. This idea was initially proposed in \cite{Asilomar2015} for a third order DPD at the IM3 sub-band, and in this paper we extend this solution to include higher order nonlinearities at the IM3 sub-band and beyond. Finally, RF measurements using practical and commercial RF equipment has been performed, thus providing a strong proof of concept for the proposed solutions.

\begin{figure*}
\centering
%\vspace{0.5cm}
\centerline{\includegraphics[width=1\textwidth]{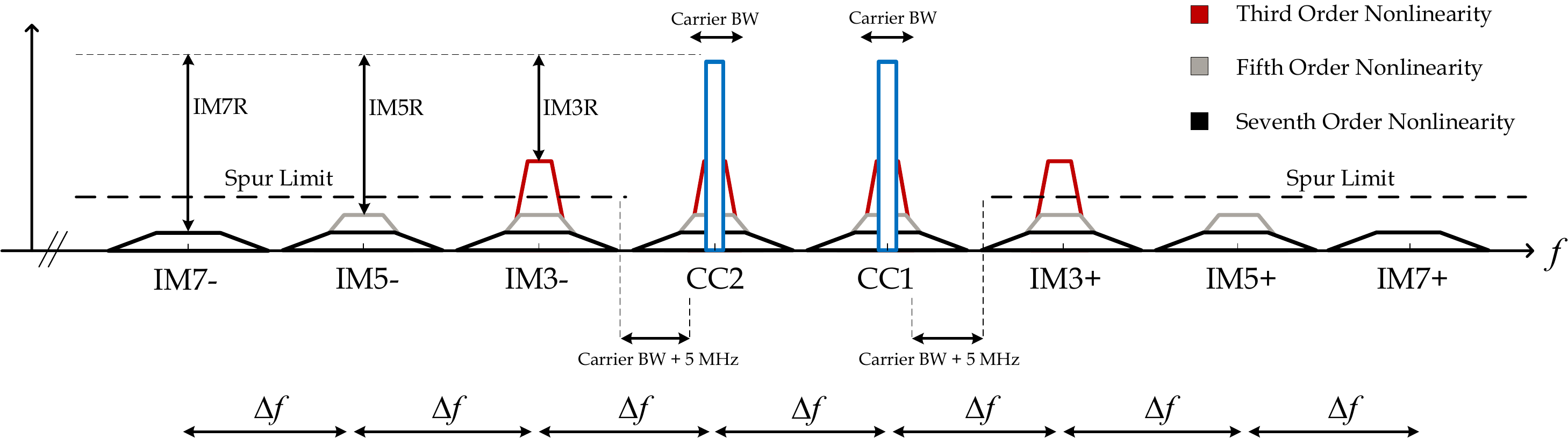}}
\caption[]{Illustration of different intermodulation distortion components created by a nonlinear PA when excited with a non-contiguous CA signal with two component carriers. Here, nonlinear distortion components up to order 7 are shown.}
\label{fig:PSD}
\end{figure*}

The rest of this article is organized as follows: Section \ref{sec:DPD_Modeling} presents the mathematical modeling of the different considered spurious components at different sub-bands, produced by a nonlinear PA with memory. Stemming from this modeling, also the corresponding core processing principles of the proposed DPD solutions at different sub-bands are formulated.
Then, Section \ref{sec:DPD_Learning} presents the proposed DPD parameter learning solutions, covering both sample-adaptive and block-adaptive decorrelation solutions at different sub-bands. Section \ref{sec:Complexity} addresses then different implementation alternatives of the proposed sub-band DPD concept, and also analyzes the computing complexity of the sub-band DPD in terms of the number of floating point operations together with some hardware complexity aspects. Finally, Sections \ref{sec:Simulations and Measurements} and \ref{sec:RF Measurement Results} report comprehensive simulation and RF measurement results, evidencing excellent spurious emission suppression in different realistic scenarios.
%tests our proposed solution on real-case simulation scenarios and practical RF measurements with commercial mobile PAs.

%Furthermore, the additional power consumed by the DPD may not justify the saving in the power amplifier efficiency for low power devices. For example, a $25\%$ power efficiency improvement due to DPD in case of a 50W (47 dBm) base station transmitter is equivalent to 12.5W power saving, while for a 0.2W (23 dBm) mobile transmitter, the same 25$\%$ improvement is equivalent to 0.05W power saving. That is why there is an urgent need to consider low complexity DPD solutions with low power consumption for mobile devices and small cells such that the power consumed by the DPD is less than the amount of power saving due to DPD. This is one of the main motivating factors for our work, where such low-power RF transmitters still suffer from lower efficiency because of the use of power back-off to control the inherent nonlinear distortion, especially in noncontiguous transmission scenarios \cite{GreenComm}.

\section{Spurious Component Modeling and Proposed Sub-band DPD Processing}
\label{sec:DPD_Modeling}

In this manuscript, we assume a practical case of non-contiguous carrier aggregation with two component carriers. When such noncontiguous dual-carrier signal is applied at the PA input, the PA nonlinearity leads to intermodulation distortion at different sub-bands, as shown in Fig. \ref{fig:PSD}. 
Assuming a CC separation of $\Delta f$, in addition to the spectral regrowth around the main carriers, intermodulation between the two CCs yields strong IMD at integer multiples of $\Delta f$ from the main CCs. 
In this article, we refer to the intermodulation (IM) sub-bands located at $\pm \Delta f$ from the main CCs as the \emph{IM3 sub-bands}. Similarly, the \emph{IM5 sub-bands} are located at $\pm 2\Delta f$ from the main CCs, and so on. 
In general, each IM sub-band includes nonlinear distortion components of different orders as shown in Fig. \ref{fig:PSD}. For example, in the case of a seventh-order PA nonlinearity, the IM3 sub-band contains third, fifth and seventh-order nonlinearities, while the IM5 sub-band contains fifth and seventh-order components.

In this section, we start with the fundamental modeling of the nonlinear distortion at the IM3 sub-band as a concrete example. Stemming from that modeling, we then formulate the basic processing of the proposed IM3 sub-band DPD. After that, the modeling is extended to cover the nonlinear distortion components at the higher-order IM sub-bands, followed by the corresponding higher-order sub-band DPD processing. The actual parameter estimation and learning algorithms for the proposed sub-band DPD structures are then presented in details in Section \ref{sec:DPD_Learning}. In all the modeling and developments, we adopt the widely-used wideband Parallel Hammerstein (PH) PA model \cite{ComplexityAnalysis} to describe the fundamental nonlinear behavior of the PA, as it has been shown to model accurately the measured nonlinear behavior of different classes of true PAs.

\subsection{Spurious Component Modeling at IM3 Sub-bands}
%In this section, we extend the third-order modeling in \cite{ICASSP2014,CROWNCOM2014} to include higher-order nonlinearities with memory effects. 
The modeling is carried out at composite baseband equivalent level, where the two component carriers are assumed to be separated by $2 f_{IF}$. The composite baseband equivalent input and output signals of the $P^{th}$ order Parallel Hammerstein PA model, with monomial nonlinearities and FIR branch filters, respectively, read

\small
\begin{align}
x(n) &= x_1(n) e^{j 2\pi \frac{f_{IF}}{f_s} n} + x_2(n) e^{-j 2\pi \frac{f_{IF}}{f_s} n}, \label{eq:PA_In} \\ 
y(n) &= \sum_{\substack{p=1 \\ p \text{ odd}}}^{P} f_{p,n} \star |x(n)|^{p-1} x(n), \label{eq:PA_out_Mem}
\end{align}
\normalsize
where $x_1(n)$ and $x_2(n)$ are the baseband component carrier signals, $f_{p,n}$ denotes the PH branch filter impulse responses of order $p$, and $\star$ is the convolution operator. Intermodulation between the two component carriers leads to the appearance of IMD components at $\pm 3 f_{IF}$, $\pm 5 f_{IF}$, etc., as shown in Fig. \ref{fig:PSD} for the corresponding RF spectrum. 

\textcolor{black}{As a concrete example, let us analyze the IMD at $\pm 3 f_{IF}$. Direct substitution of (\ref{eq:PA_In}) in (\ref{eq:PA_out_Mem}) allows extracting the baseband equivalent distortion terms at the IM3 sub-bands, which, through straight-forward manipulations, yields}

\small
\begin{align}
y_{IM3_\pm}(n) = \sum_{\substack{p=3 \\ p \text{ odd}}}^{P} f_{3\pm,p,n} \star  u_{3\pm,p}(n). \label{eq:IM3_output_Mem}
\end{align}
\normalsize
Here, $f_{3\pm,p,n}$ denote the baseband equivalent impulse responses corresponding to the wideband PH PA model filters $f_{p,n}$, evaluated at the IM3$\pm$ sub-bands, formally defined as

\small
\begin{align}
f_{3\pm,p,n} &= LPF\{e^{\mp j 2\pi \frac{3 f_{IF}}{f_s} n} f_{p,n}\}, 
\end{align}
\normalsize
with $LPF\{.\}$ denoting an ideal low pass filtering operation with a passband width $P$ times the bandwidth of the wider CC. Furthermore, $u_{3+,p}(n)$ and $u_{3-,p}(n)$ in (3) are the corresponding $p^{th}$ order static nonlinear (SNL) basis functions, related to the nonlinear distortion at the IM3+ and IM3- sub-bands, respectively. Assuming then an $11^{th}$ order PA model (i.e., $P = 11$), as a concrete high-order example, the IM3+ sub-band basis functions read

\small
\begin{align}
u_{3+,3}(n) &= x_2^*(n)x_1^2(n) \label{eq:IM3_BasisFunctions1}\\
u_{3+,5}(n) &= u_{3+,3}(n) \times (2 |x_1(n)|^2 + 3 |x_2(n)|^2)  \label{eq:IM3_BasisFunctions2}\\
u_{3+,7}(n) &= u_{3+,3}(n) \times (3 |x_1(n)|^4 + 6 |x_2(n)|^4 \nonumber\\
&+ 12 |x_1(n)|^2 |x_2(n)|^2) \label{eq:IM3_BasisFunctions3}\\
u_{3+,9}(n) &= u_{3+,3}(n) \times (4 |x_1(n)|^6 + 10|x_2(n)|^6 \nonumber\\
&+ 30 |x_1(n)|^4 |x_2(n)|^2 + 40 |x_1(n)|^2 |x_2(n)|^4) \label{eq:IM3_BasisFunctions4}\\
u_{3+,11}(n) &= u_{3+,3}(n) \times (5 |x_1(n)|^8 + 15|x_2(n)|^8 \nonumber\\
&+ 60 |x_1(n)|^6 |x_2(n)|^2 + 100|x_1(n)|^2 |x_2(n)|^6 \nonumber\\
&+ 150|x_1(n)|^4 |x_2(n)|^4). \label{eq:IM3_BasisFunctions5}
\end{align}
\normalsize
The corresponding basis functions for the IM3- sub-band, i.e., $u_{3-,p}(n)$, can be obtained by simply interchanging $x_1(n)$ and $x_2(n)$ in the above expressions. Next, these behavioral modeling results are utilized to formulate the proposed IM3 sub-band DPD concept, specifically tailored to suppress the distortion at the IM3 sub-band. 

%While the PA output signal contains also other distortion terms at higher-order IM sub-bands, we next with formulating a low-complexity sub-band DPD solution that aims at reducing these particular IM3 spurious components, followed by a similar solution for the higher-order IM sub-bands. This is formulated next at structural level in section \ref{sec:sub-band DPD principle} and section \ref{sec:higher-order DPD principle}, while the actual parameter optimization and practical estimation through the decorrelation-based principle are addressed in Section \ref{sec:DPD_Learning}.

\subsection{Proposed IM3 Sub-band DPD Principle}
\label{sec:sub-band DPD principle}
\begin{figure}
\centering
\vspace{0.5cm}
\centerline{\includegraphics[width=0.48\textwidth]{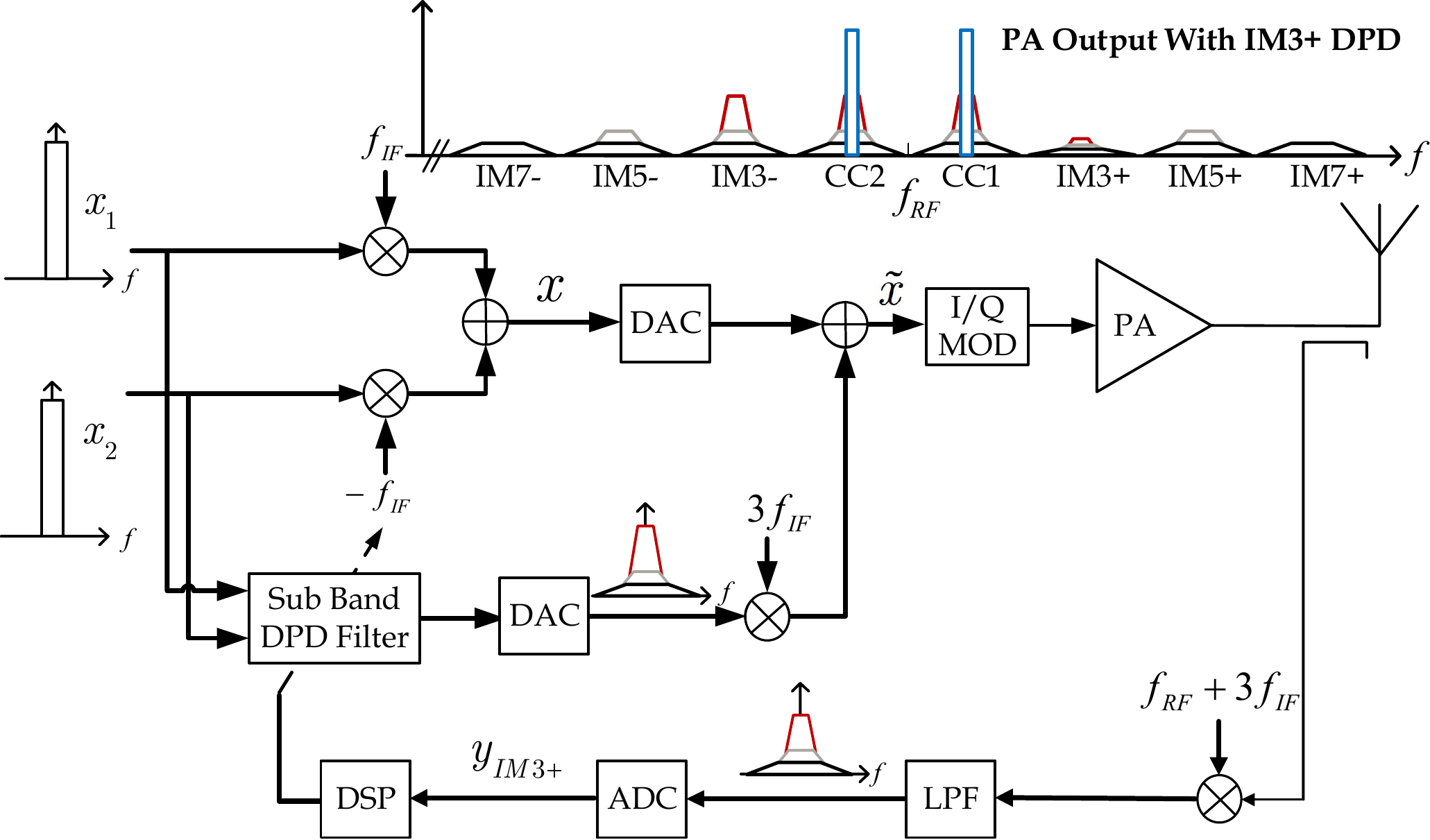}}
\caption[]{The proposed sub-band DPD processing principle focusing on the IM3+ sub-band. Thick lines are used to indicate complex I/Q processing. For presentation simplicity, TX filtering between the feedback coupler and the antenna is not shown.}
\label{fig:sub-band DPD}
\end{figure}
For presentation purposes, we focus here on suppressing the spurious emissions at the IM3+ sub-band, while the corresponding emissions at the IM3- sub-band can also be easily mitigated using a very similar structure with minor changes as will be elaborated later in Section \ref{sec:higher-order DPD principle}. The key idea of the sub-band DPD concept is to inject a proper additional low-power cancellation signal, with structural similarity to (\ref{eq:IM3_output_Mem}) and located at $+3 f_{IF}$, %when interpreted for the composite baseband equivalent signal, 
into the PA input, such that the level of the IM3+ term at the PA output is reduced. Now, stemming from the IMD structure in (\ref{eq:IM3_output_Mem}), an appropriate digital injection signal can be obtained by adopting the IM3+ basis functions $u_{3+,p}(n)$ in (\ref{eq:IM3_BasisFunctions1})-(\ref{eq:IM3_BasisFunctions5}), combined with proper filtering using a bank of sub-band DPD filters $\alpha_{3+,p,n}$. %The DPD filter coefficients for a given $p^{th}$ order IM3+ basis function thus read $\alpha_{3+,p,0}, \alpha_{3+,p,1}, ..., \alpha_{3+,p,N}$, where $N$ denotes the filter order. The selection and optimization of the filter coefficients $\alpha_{3+,p,n}$ have naturally direct impact on the amount of achievable IM3+ spur suppression, and will be addressed in detail in Section III. 
Incorporating such sub-band DPD processing with polynomial order $Q$, the composite baseband equivalent PA input signal reads

\small
\begin{align}
\tilde{x}(n) &=  x(n) + \left[\sum_{\substack{p=3 \\ p \text{ odd}}}^{Q} \alpha_{3+,p,n} \star  u_{3+,p}(n)\right]e^{j 2\pi \frac{3 f_{IF}}{f_s} n}
\label{eq:PA_In_with_IM3_DPD}
\end{align}
\normalsize
Here, and in the continuation, we use $\tilde{(.)}$ variables to indicate sub-band DPD processing and the corresponding predistorted signals.
This DPD processing principle is illustrated in Fig. \ref{fig:sub-band DPD} at a conceptual level, where the TX/RX duplexer filters are omitted for simplicity of the presentation since they do not directly impact the sub-band DPD processing or learning. Similar convention is followed also in other figures. Different implementation alternatives as well as the parameter learning and feedback receiver aspects are addressed in more details in Sections III and IV.

%Estimating the sub-band DPD filter coefficients $\alpha_{3+,p,n}$ in an optimum manner with low complexity is the main objective of this article. Theoretically speaking, the optimum value of the DPD coefficients are the ones that minimize the expected value of the power of the spurious emission in the IM3+ sub-band. It will be shown in section \ref{sec:DPD_Analytic_Solutions} that this optimum solution is computationally hungry and not easily implemented, especially in small devices. Instead, a sub-band DPD processing principle which is based on the minimization of the correlation between the baseband equivalents of the IM3+ filtered signal at the PA output $y_{IM3_+}(n)$ and the IM3+ basis functions in (\ref{eq:IM3_BasisFunctions1}-\ref{eq:IM3_BasisFunctions5}) is proposed in this article.
%Similarly, if the IM3- sub-band is targeted, the correlation between the IM3- filtered signal at the PA output and the IM3- basis functions is minimized leading to a suppression of the spurious IMD at the target sub-band.
%This idea will be explained in details in section \ref{sec:DPD_Learning}, along with a practical implementation using a simple adaptive filter learning algorithm suitable for small computationally-limited devices. 

\subsection{Generalization to Higher-Order IM Sub-bands}
\label{sec:higher-order DPD principle}
The direct substitution of (\ref{eq:PA_In}) in (\ref{eq:PA_out_Mem}) leads to the appearance of spurious intermodulation terms also at higher-order IM sub-bands, in addition to the previously considered IM3 sub-bands, as illustrated already in Fig. \ref{fig:PSD}. These higher-order sub-band emissions can also be harmful, since they can violate the emission limits, cause own receiver desensitization, or both. Thus, developing a sub-band DPD solution that can tackle also the distortion at these higher-order IM sub-bands with feasible complexity is important.
Similar to the above IM3+ sub-band developments, we next extract the IMD terms at the higher-order IM sub-bands. The baseband equivalent IMD terms at the IM5+, IM7+, IM9+, and IM11+ sub-bands, as concrete examples, can be extracted using (\ref{eq:PA_In}) and (\ref{eq:PA_out_Mem}), interpreted at proper sub-bands, yielding

\small
\begin{align}
y_{IM5_+}(n) &= \sum_{\substack{p=5 \\ p \text{ odd}}}^{P} f_{5+,p,n} \star  u_{5+,p}(n) \label{eq:IM5_output} \\ 
y_{IM7_+}(n) &= \sum_{\substack{p=7 \\ p \text{ odd}}}^{P} f_{7+,p,n} \star  u_{7+,p}(n) \\
y_{IM9_+}(n) &= \sum_{\substack{p=9 \\ p \text{ odd}}}^{P} f_{9+,p,n} \star  u_{9+,p}(n) \\
y_{IM11_+}(n) &= \sum_{\substack{p=11 \\ p \text{ odd}}}^{P} f_{11+,p,n} \star  u_{11+,p}(n). \label{eq:IM11_output}
\end{align}
\normalsize
In the above models, $f_{5+,p,n}$, $f_{7+,p,n}$, $f_{9+,p,n}$, and $f_{11+,p,n}$ denote the baseband equivalent impulse responses corresponding to the wideband PA model filters $f_{p,n}$ in (2), evaluated at the IM5+, IM7+, IM9+, and IM11+ sub-bands, respectively. These are formally obtained similar to (4) but replacing $3f_{IF}$ with either $5f_{IF}$, $7f_{IF}$, $9f_{IF}$, or $11f_{IF}$, respectively.
The $p^{th}$ order basis functions at the these higher-order IM sub-bands, denoted by $u_{5+,p}(n)$, $u_{7+,p}(n)$, $u_{9+,p}(n)$, and $u_{11+,p}(n)$, and assuming again an $11^{th}$ order PH PA model, read 

\small
\begin{align}
u_{5+,5}(n) &= (x_2(n)^*)^2 x_1(n)^3 \label{eq:IMD_BasisFunctions1} \\
u_{5+,7}(n) &= u_{5+,5}(n) (4 |x_2(n)|^2 + 3 |x_1(n)|^2) \\
u_{5+,9}(n) &= u_{5+,5}(n) (10 |x_2(n)|^4 + 6 |x_1(n)|^4 \nonumber\\
&+ 20 |x_1(n)|^2 |x_2(n)|^2) \\
u_{5+,11}(n) &= u_{5+,5}(n) (20 |x_2(n)|^6 + 10 |x_1(n)|^6 \nonumber\\
&+ 75 |x_2(n)|^4 |x_1(n)|^2 + 60 |x_2(n)|^2 |x_1(n)|^4) \\
u_{7+,7}(n) &= (x_2(n)^*)^3 x_1(n)^4 \\
u_{7+,9}(n) &= u_{7+,7}(n) (5 |x_2(n)|^2 + 4 |x_1(n)|^2) \\
u_{7+,11}(n) &= u_{7+,7}(n) (15 |x_2(n)|^4 + 10 |x_1(n)|^4 \nonumber\\
&+ 30 |x_2(n)|^2 |x_1(n)|^2) \\
u_{9+,9}(n) &= (x_2(n)^*)^4 x_1(n)^5 \\
u_{9+,11}(n) &= u_{9+,9}(n) (6 |x_2(n)|^2 + 5 |x_1(n|^2) \\ 
u_{11+,11}(n) &= (x_2(n)^*)^5 x_1(n)^6. \label{eq:IMD_BasisFunctions2}
\end{align}
\normalsize
The corresponding baseband equivalent IMD terms at the negative IM sub-bands can be obtained by simply interchanging $x_1(n)$ and $x_2(n)$ in (\ref{eq:IMD_BasisFunctions1})-(\ref{eq:IMD_BasisFunctions2}).
%the SNL basis functions expressions in (\ref{eq:IM3_BasisFunctions1})-(\ref{eq:IM3_BasisFunctions5}) and in (\ref{eq:IMD_BasisFunctions1})-(\ref{eq:IMD_BasisFunctions2}). 
%The sub-band DPD processing principle can thus also be generalized to include these higher-order IMD sub-bands using similar processing as in Fig. \ref{fig:sub-band DPD}. This will be explained in more details in the next section.

Then, stemming from the distortion structure in (\ref{eq:IM5_output})-(\ref{eq:IM11_output}), and adopting similar ideology as in the previous IM3 sub-band DPD case, a natural injection signal for suppressing the spur components at higher-order IM sub-bands can be obtained by properly filtering and combining the above higher-order sub-band basis functions. The sub-band specific filters for the $p^{th}$ order basis functions are denoted with $\alpha_{5+,p,n}$, $\alpha_{7+,p,n}$, $\alpha_{9+,p,n}$, and $\alpha_{11+,p,n}$. 
%$u_{5+,p}(n)$, $u_{7+,p}(n)$, $u_{9+,p}(n)$, and $u_{11+,p}(n)$ in (\ref{eq:IMD_BasisFunctions1}-\ref{eq:IMD_BasisFunctions2}). 
Incorporating such DPD processing with DPD polynomial order $Q$, and aggregating at the same time parallel sub-band DPDs simultaneously for the IM3+, IM5+, and IM7+ sub-bands, as a concrete example, the composite baseband equivalent PA input signal now reads

\small
\begin{align}
\tilde{x}(n) =  x(n) &+ \left[\sum_{\substack{p=3 \\ p \text{ odd}}}^{Q} \alpha_{3+,p,n} \star  u_{3+,p}(n)\right]e^{j 2\pi \frac{3 f_{IF}}{f_s} n} \nonumber\\
&+ \left[\sum_{\substack{p=5 \\ p \text{ odd}}}^{Q} \alpha_{5+,p,n} \star  u_{5+,p}(n)\right]e^{j 2\pi \frac{5 f_{IF}}{f_s} n} \nonumber\\ 
&+ \left[\sum_{\substack{p=7 \\ p \text{ odd}}}^{Q} \alpha_{7+,p,n} \star  u_{7+,p}(n)\right]e^{j 2\pi \frac{7 f_{IF}}{f_s} n} \label{eq:PA_In_DPD_IMDs}
\end{align}
\normalsize

In general, the achievable suppression of the distortion at different IM sub-bands depends directly on the selection and optimization of the different filters $\alpha_{3+,p,n}$, $\alpha_{5+,p,n}$, $\alpha_{7+,p,n}$, and so on. This is addressed in detail in the next section.
% objective, in this particular example, becomes searching for the optimum values of the filter coefficients $\alpha_{3+,p,n}$, $\alpha_{5+,p,n}$, and $\alpha_{7+,p,n}$, such that the spurious emission power at each IM sub-band is minimized. In the next section, different learning philosophies for the sub-band DPD coefficients are presented, along with a practical and simple adaptive filter implementation of the decorrelation-based algorithm, which we propose as a very nice trade-off between sub-band DPD complexity and linearization performance.

\section{Sub-band DPD Parameter Learning Using the Decorrelation Principle}
\label{sec:DPD_Learning}
In this section, based on the previous spurious component modeling and the proposed sub-band DPD principle, we formulate computing feasible and highly efficient estimation algorithms for learning and optimizing the sub-band DPD filter coefficients such that the spurious emissions are minimized at the considered IM sub-bands. We start by deriving some analytical reference solutions, taking the third-order IM3 sub-band DPD as a simple and tractable example, in order to demonstrate that minimizing the IM3 power is essentially identical to decorrelating the IM3 observation against the corresponding sub-band DPD basis functions. Then, both sample-adaptive and block-adaptive decorrelation based learning rules are devised, covering the general cases of higher-order processing with memory at IM3 and higher-order IM sub-bands.
% a practical adaptive filter implementation of the proposed decorrelation-based sub-band DPD is presented, illustrating the sample-based DPD parameter update learning rule with higher-order nonlinearities, higher-order IM sub-bands, and memory effects included in the learning algorithm. Finally, a block-adaptive version of the proposed sub-band DPD is presented as a practical solution that can tolerate loop delays inherent in real-time implementations.  

\subsection{Analytical Reference Solutions}
\label{sec:DPD_Analytic_Solutions}
In this subsection, we derive three analytical reference solutions for calculating the IM3 sub-band DPD coefficients. The considered approaches are the third-order inverse solution, also adopted in \cite{P.RoblinJan.2008}, the IM3 power minimization solution, and the analytical IM3 decorrelation-based solution. 
To keep the analytical developments simple and tractable, we consider a simplified case with a third-order memoryless IM3+ sub-band DPD and a third-order memoryless PA. The actual sample-adaptive and block-adaptive decorrelation based learning solutions, devised later in this section, are formulated for the general cases of higher-order processing with memory.

Now, starting with the dual carrier signal in (\ref{eq:PA_In}), and limiting the study to the simplified case of a third-order memoryless IM3+ sub-band DPD and a third-order memoryless PA, the basic signal models are given by (\ref{eq:ThirdOrderAnalysis_1})-(\ref{eq:ThirdOrderAnalysis}).
%in (\ref{eq:ThirdOrderAnalysis_1})-(\ref{eq:ThirdOrderAnalysis}).

\small
\begin{align}
y(n) &= f_1 x(n) + f_3 |x(n)|^2 x(n), \label{eq:ThirdOrderAnalysis_1} \\
u(n) &= x_2^*(n)x_1^2(n), \\
y_{IM3_+}(n) &= f_3 u(n), \\
\tilde{x}(n) &=  x(n) + \alpha u(n) e^{j 2\pi \frac{3 f_{IF}}{f_s} n}, \\
\tilde{y}_{IM3_+}(n) &=  (f_3 + f_1 \alpha) u(n) + 2 f_3 \alpha (|x_1(n)|^2 + |x_2(n)|^2) u(n) \nonumber\\
&+ f_3 |\alpha|^2 \alpha |x_1(n)|^4 |x_2(n)|^2 u(n).
\label{eq:ThirdOrderAnalysis}
\end{align}
\normalsize
Here, $f_1$ and $f_3$ are the memoryless PA model parameters, $\alpha$ denotes the memoryless sub-band DPD parameter to be optimized, while ${y}_{IM3_+}(n)$ and $\tilde{y}_{IM3_+}(n)$ refer to the baseband equivalent PA output at the positive IM3 sub-band without and with sub-band DPD, respectively.

Now, as (\ref{eq:ThirdOrderAnalysis}) clearly shows, the IM3 sub-band distortion at the PA output, with DPD adopted, depends directly on, and can thus be controlled by, the DPD coefficient $\alpha$. 
In the well-known \emph{third-order inverse solution}, the DPD parameter $\alpha$ is selected such that the third-order term in (\ref{eq:ThirdOrderAnalysis}) is canceled, i.e., $f_3 + f_1 \alpha = 0$. The corresponding solution, denoted by $\alpha_{inv}$, thus reads

\small
\begin{align}
\alpha_{inv} &= -f_3 / f_1.  \label{eq:3rd_order_inv}
\end{align}
\normalsize
However, this will not remove all the distortion, because higher-order terms will be created due to the predistortion, as shown in (\ref{eq:ThirdOrderAnalysis}).

A more elaborate method for selecting the DPD parameter $\alpha$ is the one that minimizes the power of the total IM3 sub-band signal, $\tilde{y}_{IM3_+}(n)$, referred to as the \emph{minimum IM3 power} or \emph{minimum mean-squared error (MMSE)} solution in the continuation. For notational convenience, we define the so-called error signal as $e(n)=\tilde{y}_{IM3_+}(n)$, as with ideal predistortion the IM3 sub-band signal would be zero, and thus the optimization means minimizing the power of this error signal. 
%Thus, the solution can be considered the minimum mean squared error (MMSE) solution $\alpha_{MMSE}$. 
%From (\ref{eq:ThirdOrderAnalysis}), the error signal $e(n)$ reads  
%\small
%\begin{align}
%e(n) &= \tilde{y}_{IM3_+}(n) \nonumber\\
%&= (\beta_3 + \beta_1 \alpha) u(n) + 2 \beta_3 \alpha (|x_1(n)|^2 + |x_2(n)|^2) u(n) \nonumber\\
%&+ \beta_3 |\alpha|^2 \alpha |x_1(n)|^4 |x_2(n)|^2 u(n)
%\label{eq:Error_Signal}
%\end{align}
%\normalsize
The detailed derivation of the optimum DPD parameter which minimizes the mean squared error $\E[|e(n)|^2]$ is given in Appendix A, yielding
%$\E[e(n)e^*(n)]$

\small 
\begin{align}
&\alpha_{MMSE} = \nonumber\\
&\dfrac{-\left[f_1 f_3^* \E_{42} + 2 |f_3|^2 (\E_{62} + \E_{44})\right]^*}{\left[|f_1|^2 \E_{42} + 4 \mathbb{R}(f_1 f_3^*)(\E_{62} + \E_{44}) 
+ 4 |f_3|^2 (\E_{46} + 2 \E_{64} + \E_{82})\right]^*}, \label{eq:MMSE}
\end{align}
\normalsize
where $\E_{ij}$ refers to the products of the CC signals' higher-order moments of the form $\E[|x_1|^i] \E[|x_2|^j]$, while $\E[.]$ denotes the statistical expectation operator. 

As will be shown with concrete examples in Section V, this MMSE solution provides better linearization performance compared to the third-order inverse solution. 
However, as shown in (\ref{eq:MMSE}), the analytical MMSE solution requires the knowledge of various higher-order moments of the CC signals. Furthermore, the above solution is valid only in the case of a memoryless third-order nonlinear system, beyond which obtaining an analytical expression for the DPD coefficients becomes overly tedious. \textcolor{black}{Thus, to relax these constraints, an alternative solution, based on minimizing the correlation between the IM3 sub-band observation and the third-order basis function $u(n) = x_2^*(n)x_1^2(n)$ was initially discussed in \cite{ICASSP2014}, and will be extended to higher nonlinearity orders and higher-order IM sub-bands in this paper}. The analytical reference solution for the decorrelation-based sub-band DPD structure is obtained by setting the correlation between the error signal, $e(n)$, and the third-order basis function, $u(n) = x_2^*(n)x_1^2(n)$, to zero, i.e., $\E[e(n)u^*(n)] = 0$. Then, through straight-forward algebraic manipulations, the decorrelation-based DPD coefficient, denoted by $\alpha_{o}$, can be shown to read

\small
\begin{align}
\alpha_{o} = \frac{-f_3}{f_1 + 2 f_3 (\frac{\E_{60}}{\E_{40}} + \frac{\E_{04}}{\E_{02}})}. \label{eq:alpha_0}
%\alpha_{o} &= \frac{-f_3 \E[|x_1|^4] \E[|x_2|^2]}{f_1 \E[|x_1|^4] \E[|x_2|^2] + 2 f_3 (\E[|x_1|^6] \E[|x_2|^2] + \E[|x_1|^4] \E[|x_2|^4])} \nonumber\\
%&= \frac{-f_3}{f_1 + 2 f_3 (\frac{\E_{60}}{\E_{40}} + \frac{\E_{04}}{\E_{02}})} \label{eq:alpha_0}
\end{align} 
\normalsize
Assuming then, for simplicity, that the CC baseband equivalents are complex Gaussians, (\ref{eq:alpha_0}) simplifies to

\small
\begin{align}
\alpha_{o}= \frac{-f_3}{f_1 + 2 f_3 (3 \sigma^2_{x,1} + 2 \sigma^2_{x,2})}, 
\end{align}
\normalsize
where $\sigma^2_{x,1} = \E[|x_1|^2]$ and $\sigma^2_{x,2} = \E[|x_2|]^2$ are the variances of the two CCs \cite{ICASSP2014}.

A clear advantage of the decorrelation-based approach compared to the earlier MMSE solution lies in the simple and straightforward adaptive filtering based practical computing solutions, sketched initially in \cite{ICASSP2014} for simple third-order processing at IM3, which do not require any prior knowledge about signal moments or any other parameters. %, that does not require any higher order moment evaluations as in the MMSE solution. This adaptive learning structure will then also enable directly tracking, e.g., possible time-variations in the PA characteristics due to temperature changes and other possible sources like device ageing. 
Furthermore, opposed to the MMSE solution, the decorrelation based approach can be easily extended to include higher-order nonlinearities and memory effects, at both the IM3 sub-band as well as the other higher-order IM sub-bands, as will be described in details in the next subsection.

%The decorrelation-based and MMSE DPD solutions are compared using Monte-Carlo simulations in section \ref{sec:Analytical_Sim} and are shown to give almost exactly the same performance, thus justifying the decorrelation-based solution as a near optimum solution in the sense of minimizing the power in the IM3 sub-band. 

\subsection{Sample-Adaptive Decorrelation-based Learning}
\label{sec:DPD_adaptive_filter}
\begin{figure*}[t!]
\centering
%\vspace{0.5cm}
\centerline{\includegraphics[width=0.95\linewidth]{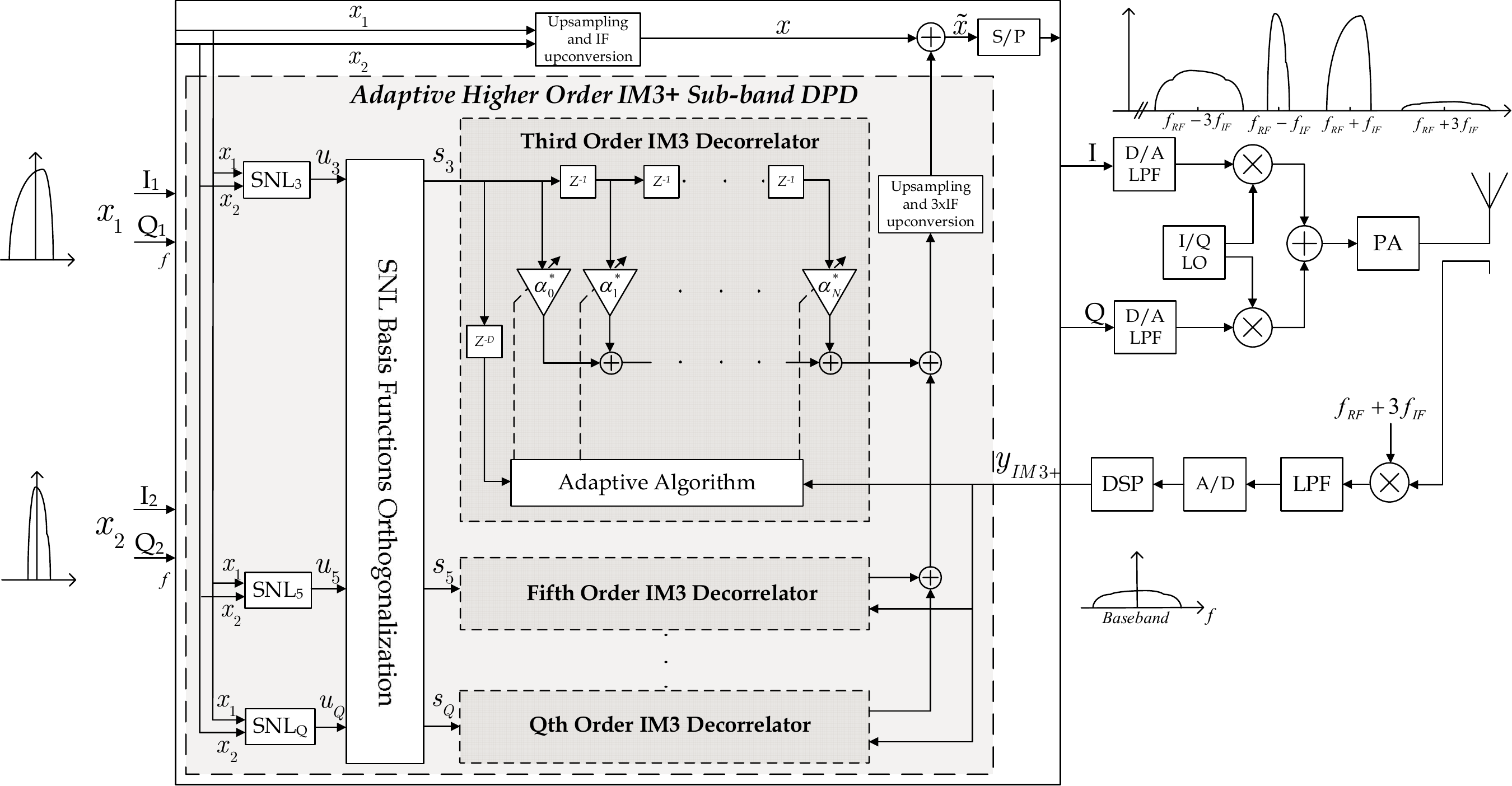}}
\caption[]{Detailed block diagram of the sample-adaptive decorrelation-based IM3+ sub-band DPD with higher-order nonlinearities and memory.}
\label{fig:Decorr_DPD}
\end{figure*}

In this subsection, we provide the actual sample-adaptive learning rules for both IM3 and higher-order IM sub-band DPD structures with memory, whose basic operating principles were described in Section II. To facilitate the learning, we assume a feedback or observation receiver measuring the particular IM sub-band whose sub-band DPD filter coefficients are currently under learning. Notice that opposed to classical wideband DPD principles, where a wideband observation receiver is typically needed, a more narrowband receiver is sufficient here since only the particular IM sub-band is observed. As both the IM3 and the higher-order IM sub-band DPDs are based on multiple strongly correlated basis functions, given in (\ref{eq:IM3_BasisFunctions1})-(\ref{eq:IM3_BasisFunctions5}) and (\ref{eq:IMD_BasisFunctions1})-(\ref{eq:IMD_BasisFunctions2}), respectively, we start by introducing a basis function orthogonalization procedure, after which the actual proposed adaptive decorrelation algorithms are described.

% the analysis was based on third-order memoryless processing, which we extend in this subsection to higher order nonlinearities with memory effects.  In order to make the presentation tractable, we first start with developing the DPD learning rule for the IM3+ sub-band DPD coefficients, after which we extend the learning algorithm to include other IM sub-bands in a straight forward manner.
%As previously explained in section \ref{sec:sub-band DPD principle}, the PA output at the IM3+ sub-band in (\ref{eq:IM3_output_Mem}) includes delayed replicas of the IM3+ SNL basis functions.
%This property shall be used in our design proposal for linearizing PA's exhibiting memory effects. The key idea is to inject a proper additional low-power cancellation signal, similar in form to (\ref{eq:IM3_output_Mem}) and located at $+3 f_{IF}$, into the PA input signal (\ref{eq:PA_In}), such that the level of the IM3+ term at the PA output is reduced. The main principle of our proposed decorrelation-based learning algorithm is to select the DPD coefficients $\alpha_{3+,p,n}$ in (\ref{eq:PA_In_with_IM3_DPD}) that minimize the correlation between the IM3+ SNL basis functions in (\ref{eq:IM3_BasisFunctions1} - \ref{eq:IM3_BasisFunctions5}), and the PA output at the IM3+ sub-band.

\subsubsection{Basis Function Orthogonalization}
When a nonlinearity order higher than the IM sub-band order is considered, %higher-order nonlinearities are considered already at the IM3 sub-band, 
multiple basis functions are adopted in the sub-band DPD processing, as described in Section II. % in the adaptive learning algorithm compared to third-order nonlinear processing in \cite{ICASSP2014}, which included only one third-order basis function in the IM3+ sub-band. The challenge comes from the fact that 
Taking as an example case the IM3+ sub-band, the SNL basis functions $u_{3+,p}(n)$, $p = 1, 3, ..., Q$, given in (\ref{eq:IM3_BasisFunctions1})-(\ref{eq:IM3_BasisFunctions5}), are highly correlated. This can negatively impact the convergence and stability of the adaptive decorrelation-based learning. Therefore, the SNL basis functions are first orthogonalized \cite{OrthPloyDPD}, which yields a new set of DPD basis functions $s_{3+,p}(n)$, $p = 1, 3, ..., Q$, written formally at sample level as 
%with nonlinearity order $Q$ in this particular example of the IM3+ sub-band, are thus obtained from the old set using a simple linear transformation, as

\small
\begin{align}
\textbf{s}_{3+}(n) &= \textbf{W} \textbf{u}_{3+}(n),
\label{eq:QR}
\end{align}
\normalsize
where

\small
\begin{align}
\textbf{u}_{3+}(n) &= [u_{3+,3}(n)\:\:u_{3+,5}(n) \:\: ... \:\: u_{3+,Q}(n)]^T, \\
\textbf{s}_{3+}(n) &= [s_{3+,3}(n)\:\:s_{3+,5}(n) \:\: ... \:\: s_{3+,Q}(n)]^T, \\
\textbf{W} &= \begin{bmatrix}
    1 & 0 & 0 & 0  & 0 \\
    w_{5,3} & w_{5,5} & 0       & 0      & 0 \\
	  w_{7,3} & w_{7,5} & w_{7,7} & \dots  & 0 \\
		\vdots  & \vdots  & \vdots  & \ddots & \vdots \\   		
    w_{Q,3} & w_{Q,5} & w_{Q,7} & \dots  & w_{Q,Q} 
\end{bmatrix}.
\end{align}
\normalsize
The lower triangular matrix $\textbf{W}$ can be obtained, e.g., through Gram-Schmidt orthogonalization, QR or singular value decomposition, or using a lower complexity iterative orthogonalization algorithm \cite{Iterative_Orth}. 

%{\color{red} QUESTION: should we show extension to higher-order IM sub-band, or at least comment it shortly with a sentence or two ? }
Similarly, the SNL basis functions in (\ref{eq:IMD_BasisFunctions1})-(\ref{eq:IMD_BasisFunctions2}) corresponding to the higher-order IM sub-bands are also orthogonalized, to obtain new sets of orthogonal basis functions $\textbf{s}_{5+}(n)$, $\textbf{s}_{7+}(n)$, etc.
%To conclude, the IM3+ sub-band DPD main-path processing goes as follows: the basis functions $u_{3+,p}(n)$ are first orthogonalized as in (\ref{eq:QR}) to obtain $s_{3+,p}(n)$, then each of these new orthogonal basis functions is filtered with an $N$ tap filter $\alpha_{3+,p,n}$ before being added to the PA input at the corresponding IM3+ sub-band frequency as illustrated also in Fig. \ref{fig:Decorr_DPD}. 

\subsubsection{Adaptive Learning for IM3+ Sub-band}
We now present the actual sample-adaptive decorrelation based learning algorithm for the IM3+ sub-band DPD coefficients. For notational convenience, we introduce the following vectors

\small
\begin{align}
\boldsymbol{\alpha}_{3+,l}(n) &= [\alpha_{3+,3,l}(n)\:\:\alpha_{3+,5,l}(n)\:\: ... \:\: \alpha_{3+,Q,l}(n)]^T, \\
\bar{\boldsymbol{\alpha}}_{3+}(n) &= [\boldsymbol{\alpha}_{3+,0}(n)^T \:\: \boldsymbol{\alpha}_{3+,1}(n)^T \:\: ... \:\: \boldsymbol{\alpha}_{3+,N}(n)^T]^T, \\
\bar{\textbf{s}}_{3+}(n) &= [\textbf{s}_{3+}(n)^T \:\: \textbf{s}_{3+}(n-1)^T \:\: ... \:\: \textbf{s}_{3+}(n-N)^T]^T,
\end{align}
\normalsize 
where now $\alpha_{3+,p,l}(n)$ denotes the $l^{th}$ adaptive filter coefficient of the $p^{th}$ order orthogonalized IM3+ basis function $s_{3+,p}(n)$ at time index $n$, and $N$ denotes the adaptive filter memory depth. Furthermore, the vectors $\bar{\boldsymbol{\alpha}}_{3+}(n)$ and $\bar{\textbf{s}}_{3+}(n)$ incorporate all the coefficients and basis function samples up to polynomial order $Q$. Adopting this notation, the instantaneous sample of the composite baseband equivalent PA input signal $\tilde{x}(n)$ now reads

\small
\begin{align}
\tilde{x}(n) &=  x(n) + \tilde{x}_{3+}(n)e^{j 2\pi \frac{3 f_{IF}}{f_s} n},
\end{align}
\normalsize
where the instantaneous sample of the baseband equivalent IM3+ injection signal $\tilde{x}_{3+}(n)$ reads

\small
\begin{align}
\tilde{x}_{3+}(n) &= \bar{\boldsymbol{\alpha}}_{3+}(n)^H \bar{\textbf{s}}_{3+}(n). \label{eq:injected_im3}   
\end{align}
\normalsize

Then, in order to adaptively update the filter coefficients $\bar{\boldsymbol{\alpha}}_{3+}(n)$, the IM3+ sub-band is observed with the feedback receiver, and the coefficient is updated as 

\small
\begin{align}
\bar{\boldsymbol{\alpha}}_{3+}(n+1) = \bar{\boldsymbol{\alpha}}_{3+}(n) - \frac{\mu}{||\bar{\textbf{s}}_{3+}(n)||^2 + C} \: \bar{\textbf{s}}_{3+}(n)e_{3+}^*(n), 
\label{eq:DPD_FilterUpdate_Mem}
\end{align}
\normalsize
where $e_{3+}(n) = \tilde{y}_{IM3_+}(n)$ denotes the baseband equivalent observation of the PA output at the IM3+ sub-band with the current DPD coefficients. The scaling factor $||\bar{\textbf{s}}_{3+}(n)||^2 + C$ normalizing the learning step-size $\mu$ in (\ref{eq:DPD_FilterUpdate_Mem}) is philosophically similar to that of the Normalized Least Mean Square (NLMS) algorithm, in effect making the learning characteristics more robust against the input data dynamics. % factor and the constant C is added to avoid numerical problems when the energy in the current set of samples is very small. 

The proposed coefficient update in (\ref{eq:DPD_FilterUpdate_Mem}) is seeking to decorrelate the IM3+ observation and the adopted orthogonalized basis functions. This type of learning algorithm can also be interpreted as a stochastic Newton root search in the objective function  $J(\bar{\boldsymbol{\alpha}}_{3+}) = \E[\tilde{x}_{3+}(n)e_{3+}^*(n)]$, where the target is to search for the DPD coefficients $\bar{\boldsymbol{\alpha}}_{3+}$ that minimize the ensemble correlation between $e_{3+}(n)$ and the baseband equivalent of the IM3+ injection signal $\tilde{x}_{3+}(n)$ in (\ref{eq:injected_im3}). The overall processing flow is graphically illustrated in Fig. \ref{fig:Decorr_DPD}.
%Notice that the learning and the actual main DPD path filtering are separated, to be able to handle the delay of the predistortion feedback loop. This delay is denoted by the block $z^{-D}$ in front of the 'Adaptive Algorithm' block as shown in Fig. \ref{fig:Decorr_DPD}.

\subsubsection{Adaptive Learning for Higher-Order IM Sub-bands}
We now extend the decorrelation based learning to higher-order IM sub-bands. First, adopting vector-based notations again, we introduce the following vectors for notational convenience

\small
\begin{align}
\boldsymbol{\alpha}_{5+,l}(n) &= [\alpha_{5+,5,l}(n)\:\:\alpha_{5+,7,l}(n)\:\: ... \:\: \alpha_{5+,Q,l}(n)]^T, \\
\boldsymbol{\alpha}_{7+,l}(n) &= [\alpha_{7+,7,l}(n)\:\:\alpha_{7+,9,l}(n)\:\: ... \:\: \alpha_{7+,Q,l}(n)]^T, \\
\bar{\boldsymbol{\alpha}}_{5+}(n) &= [\boldsymbol{\alpha}_{5+,0}(n)^T \:\: \boldsymbol{\alpha}_{5+,1}(n)^T \:\: ... \:\: \boldsymbol{\alpha}_{5+,N}(n)^T]^T, \\
\bar{\boldsymbol{\alpha}}_{7+}(n) &= [\boldsymbol{\alpha}_{7+,0}(n)^T \:\: \boldsymbol{\alpha}_{7+,1}(n)^T \:\: ... \:\: \boldsymbol{\alpha}_{7+,N}(n)^T]^T, \\
\textbf{s}_{5+}(n) &= [s_{5+,5}(n)\:\:s_{5+,7}(n) \:\: ... \:\: s_{5+,Q}(n)]^T, \\
\textbf{s}_{7+}(n) &= [s_{7+,7}(n)\:\:s_{7+,9}(n) \:\: ... \:\: s_{7+,Q}(n)]^T, \\
\bar{\textbf{s}}_{5+}(n) &= [\textbf{s}_{5+}(n)^T \:\: \textbf{s}_{5+}(n-1)^T \:\: ... \:\: \textbf{s}_{5+}(n-N)^T]^T, \\
\bar{\textbf{s}}_{7+}(n) &= [\textbf{s}_{7+}(n)^T \:\: \textbf{s}_{7+}(n-1)^T \:\: ... \:\: \textbf{s}_{7+}(n-N)^T]^T.
\end{align}
\normalsize
Then, similar to $\tilde{x}_{3+}(n)$ in (\ref{eq:injected_im3}), the baseband equivalents of the IM5+ and IM7+ sub-band DPD injection signals, denoted by  $\tilde{x}_{5+}(n)$ and $\tilde{x}_{7+}(n)$, respectively, read

\small
\begin{align}
\tilde{x}_{5+}(n) = \bar{\boldsymbol{\alpha}}_{5+}(n)^H \bar{\textbf{s}}_{5+}(n), 
%\\
\:\:   
\tilde{x}_{7+}(n) = \bar{\boldsymbol{\alpha}}_{7+}(n)^H \bar{\textbf{s}}_{7+}(n). 
\end{align}
\normalsize
Thus, the composite baseband equivalent PA input signal when the IM3+, IM5+, and IM7+ sub-band DPDs are all included, reads

\small
\begin{align}
\tilde{x}(n) = x(n) &+ \tilde{x}_{3+}(n)e^{j 2\pi \frac{3 f_{IF}}{f_s} n} + \tilde{x}_{5+}(n)e^{j 2\pi \frac{5 f_{IF}}{f_s} n} \nonumber\\
&+ \tilde{x}_{7+}(n)e^{j 2\pi \frac{7 f_{IF}}{f_s} n}.
\end{align}
\normalsize

%\small
%\begin{align}
%\tilde{x}(n) =  x(n) &+ \left[\sum_{\substack{p=3 \\ p \text{ odd}}}^{Q} \alpha_{3+,p,n} \star  s_{3+,p}(n)\right]e^{j 2\pi \frac{3 f_{IF}}{f_s} n} \nonumber\\
%&+ \left[\sum_{\substack{p=5 \\ p \text{ odd}}}^{Q} \alpha_{5+,p,n} \star  s_{5+,p}(n)\right]e^{j 2\pi \frac{5 f_{IF}}{f_s} n} \nonumber\\ 
%&+ \left[\sum_{\substack{p=7 \\ p \text{ odd}}}^{Q} \alpha_{7+,p,n} \star  s_{7+,p}(n)\right]e^{j 2\pi \frac{7 f_{IF}}{f_s} n} \label{eq:PA_In_DPD_IMDs}
%\end{align}
%\normalsize
Then, similar to the IM3+ coefficient update in (\ref{eq:DPD_FilterUpdate_Mem}), the higher-order IM5+ and IM7+ coefficient updates read

\small
\begin{align}
\bar{\boldsymbol{\alpha}}_{5+}(n+1) &= \bar{\boldsymbol{\alpha}}_{5+}(n) - \frac{\mu}{||\bar{\textbf{s}}_{5+}(n)||^2 + C} \: \bar{\textbf{s}}_{5+}(n)e_{5+}^*(n), \\
\bar{\boldsymbol{\alpha}}_{7+}(n+1) &= \bar{\boldsymbol{\alpha}}_{7+}(n) - \frac{\mu}{||\bar{\textbf{s}}_{7+}(n)||^2 + C} \: \bar{\textbf{s}}_{7+}(n)e_{7+}^*(n), 
\end{align}
\normalsize
where $e_{5+}(n) = \tilde{y}_{IM5_+}(n)$ and $e_{7+}(n) = \tilde{y}_{IM7_+}(n)$ denote the baseband equivalents of the PA output at the IM5+ and IM7+ sub-bands, with the corresponding sub-band DPDs included adopting the current coefficients, $\bar{\boldsymbol{\alpha}}_{5+}(n)$ and $\bar{\boldsymbol{\alpha}}_{7+}(n)$, respectively. From the learning perspective, observing a single IM sub-band at a time is the most obvious alternative, which means that the learning for multiple sub-band DPDs happens one at a time. Furthermore, extending the decorrelation-based learning for the negative IM sub-bands is straight-forward. This can be obtained by interchanging $x_1(n)$ and $x_2(n)$ in the SNL basis functions expressions, and observing the PA output at the corresponding negative IM sub-bands. 

\subsection{Block-Adaptive Decorrelation-based Learning}
\label{sec:block_based_DPD}
As Fig. \ref{fig:Decorr_DPD} illustrates, the previous sample-adaptive decorrelation-based learning concept is in principle a closed-loop feedback system with nonlinear adaptive processing inside the loop. During the DPD learning phase, and under the potential hardware processing and latency constraints, the DPD parameter convergence and consequently the DPD linearization performance can be affected, especially if the learning loop delay becomes large. Stemming from this, an alternative and new block-adaptive decorrelation-based learning solution is developed next.
%The developed solution is as an extension of the third order solution presented in \cite{Asilomar2015}, where higher-order nonlinearities and higher-order IM sub-bands are included instead of only third order IM3 sub-band processing.

\begin{figure}[t!]
\centering
%\vspace{0.5cm}
\centerline{\includegraphics[width=0.95\linewidth]{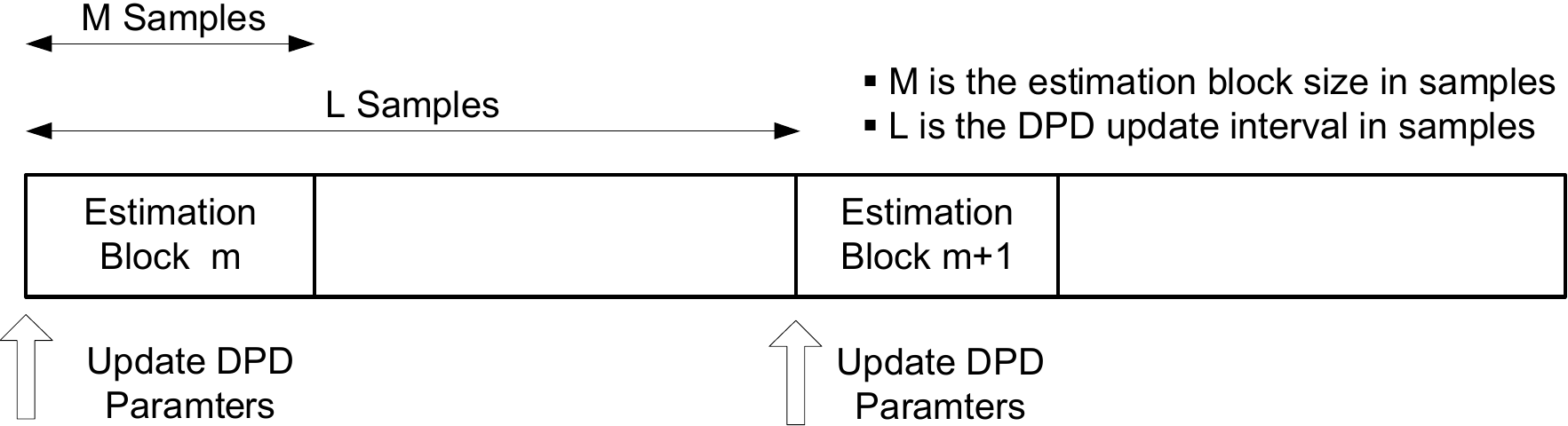}}
\caption[]{Block-adaptive DPD learning concept. The DPD parameters estimated in the current estimation block $m$ are applied in the sub-band DPD main path processing during the next block.}
\label{fig:BlockDPD_Learning}
\end{figure}

The proposed block-adaptive learning rule implies defining two distinct blocks in the processing, as illustrated in Fig. \ref{fig:BlockDPD_Learning}. 
A single update cycle of the learning algorithm will utilize $M$ samples %, as shown in (\ref{eq:BlockAdaptive}), 
whereas the DPD parameter update interval is $L$ samples, with $M \leq L$. 
% Consequently, a loop delay up to at least the difference between the two block sizes can be tolerated for a certain DPD linearization performance. 
Thus, by proper choice of $M$ and $L$, arbitrarily long loop delays can in principle be tolerated, facilitating stable operation under various hardware and software processing latency constraints. Notice that proper timing synchronization between the observation receiver output and the basis functions is, in general, needed, which can be accomplished prior to executing the actual coefficient learning procedure.

\begin{figure}[t!]
%\centering
\subfigure[Sub-band DPD Architecture I: Digital IF injection]{
\includegraphics[width=0.5\textwidth]{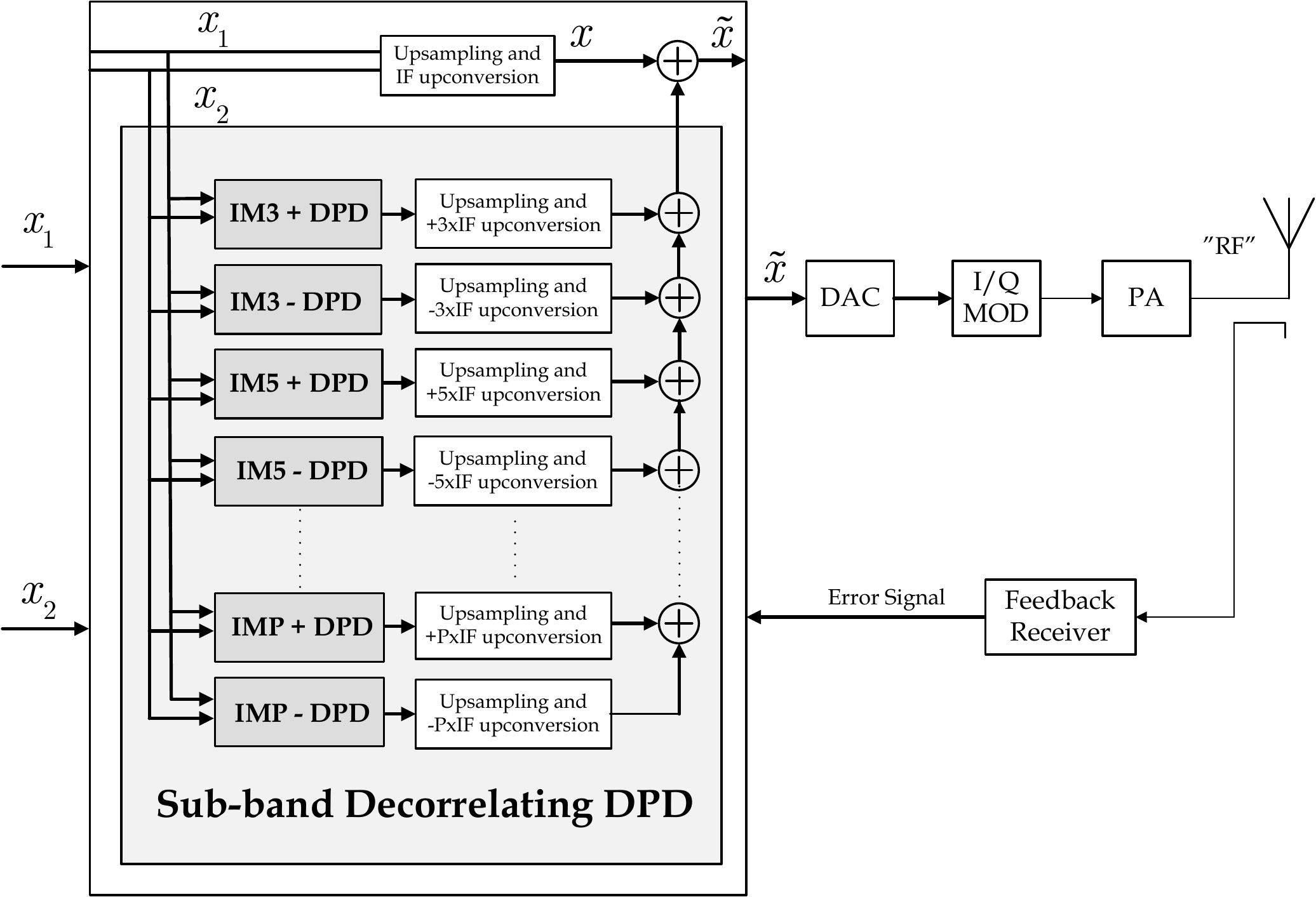}
\label{fig:Decorr_DPD_IM5_IM7_IM9}
}
\hfill
\subfigure[Sub-band DPD Architecture II: Analog IF injection]{
\includegraphics[width=0.5\textwidth]{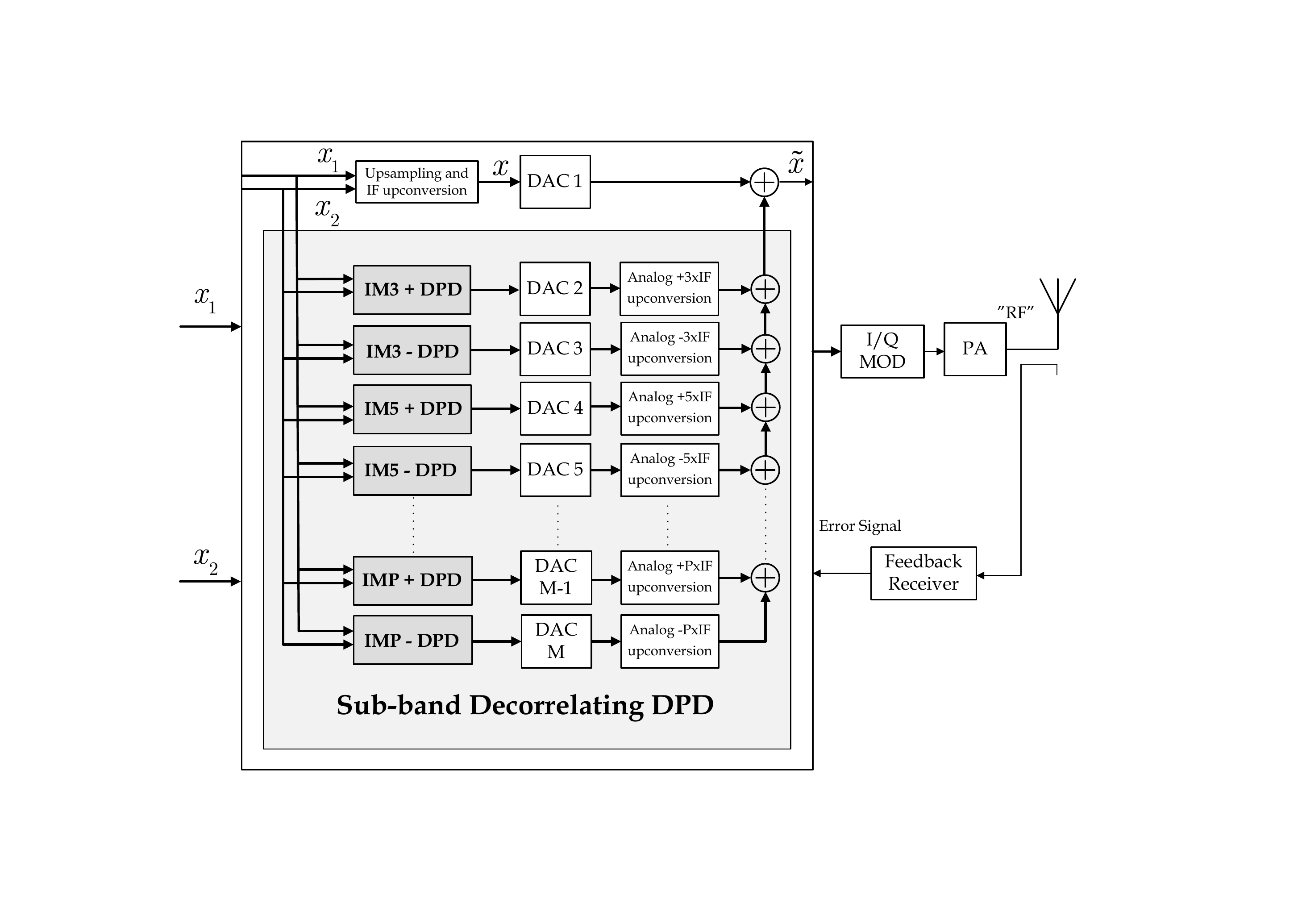}
\label{fig:Decorr_DPD_IM5_IM7_IM9_2}
}
\caption{Overall sub-band DPD architecture with multiple IM sub-bands included. Thick lines indicate complex I/Q processing.} %The architecture can use a single feedback receiver for all IM sub-bands and perform the linearization sequentially. Thick lines indicate complex I/Q processing.}
\label{fig:ImplementationAlternatives}
\end{figure}

\begin{table*}
\caption{Comparison of running complexities of ninth order sub-band and full-band DPDs. Two 1 MHz CCs with 20 MHz carrier spacing is assumed. DPD sample rate is 189 MSPS for the full-band and 9 MSPS for the sub-band DPD.}
\centering
{\begin{tabular}{lllll|l}\hline
				& \multicolumn{4}{c}{\textbf{Sub-band DPD}}      &  \multicolumn{1}{c}{\textbf{Full-band DPD}}    \\\hline
                                          & IM3+/- Sub-band  & IM5+/- Sub-band      & IM7+/- Sub-band      & IM9+/- Sub-band     & Full-band        \\\hline
\textbf{Basis Function Generation FLOPs}  & 37               & 40                   & 45           			   & 48                  & 11               \\\hline
\textbf{DPD Filtering FLOPs}  			      & 32($N_1$+1) - 2  & 24($N_1$+1) - 2      & 16($N_1$+1) - 2      & 8($N_1$+1) - 2      & 40($N_2$+1) - 2  \\\hline
\textbf{Total Number of FLOPs}				    & 32($N_1$+1) + 35 & 24($N_1$+1) + 38     & 16($N_1$+1) + 43     & 8($N_1$+1) + 46 		 & 40($N_2$+1) + 9  \\\hline
\rowcolor{Gray}
\textbf{GFLOPS ($N_1 = 1$, $N_2 = 3$)}    & \textbf{0.891}   & \textbf{0.774}       & \textbf{0.675}       & \textbf{0.558}      & \textbf{31.941}  \\\hline  
\end{tabular}}{}
\label{tab:FLOPS}
\end{table*}

Now, assuming an estimation block size of $M$ samples and DPD filter memory depth of $N$ per each of the IM3+ orthogonalized basis functions, the following vectors and matrices, which stack all the samples and the corresponding DPD filter coefficients within block $m$, can be defined:

\small
\begin{align}
\textbf{s}_{3+}(n_m) &= [s_{3+,3}(n_m)\:\:s_{3+,5}(n_m) \:\: ... \:\: s_{3+,Q}(n_m)]^T, \\
\bar{\textbf{s}}_{3+}(n_m) &= [\textbf{s}_{3+}(n_m)^T \:\: ... \:\: \textbf{s}_{3+}(n_m-N)^T]^T, \\
\textbf{S}_{3+}(m) &= [\bar{\textbf{s}}_{3+}(n_m)\:\: ... \:\: \bar{\textbf{s}}_{3+}(n_m+M-1)], \\
\boldsymbol{\alpha}_{3+,l}(m) &= [\alpha_{3+,3,l}(m)\:\:\alpha_{3+,5,l}(m)\:\: ... \:\: \alpha_{3+,Q,l}(m)]^T, \\
\bar{\boldsymbol{\alpha}}_{3+}(m) &= [\boldsymbol{\alpha}_{3+,0}(m)^T \:\: \boldsymbol{\alpha}_{3+,1}(m)^T \:\: ... \:\: \boldsymbol{\alpha}_{3+,N}(m)^T]^T.
\end{align}
\normalsize
Here, the index of the first sample of block $m$ is denoted by $n_m$. The block-adaptive IM3+ sub-band DPD coefficient update then reads

\small
\begin{align}
\textbf{e}_{3+}(m) &= [\tilde{y}_{IM3_+}(n_m) \:\: ... \:\: \tilde{y}_{IM3_+}(n_m+M-1)]^T, \\
\bar{\boldsymbol{\alpha}}_{3+}(m+1) &= \bar{\boldsymbol{\alpha}}_{3+}(m) - \frac{\mu}{||\textbf{S}_{3+}(m)||^2 + C} \: \textbf{S}_{3+}(m)\textbf{e}_{3+}^*(m), 
\label{eq:BlockAdaptive}
\end{align}
\normalsize
where $\textbf{e}_{3+}^*(m)$ refers to the element-wise conjugated error signal vector, while $\textbf{S}_{3+}(m)$ denotes the filter input data matrix, all within the processing block $m$. The obtained new DPD coefficients $\bar{\boldsymbol{\alpha}}_{3+}(m+1)$ are then applied on the next block of $L$ samples as illustrated in Fig. \ref{fig:BlockDPD_Learning}.
%The baseband equivalent of the block-based IM3+ injected signal at the PA input is thus obtained as follows
%\small
%\begin{align}
%\textbf{x}(m) &= [x(n_{m})\:\: ... \:\: x(n_{m}+L-1)]^T, \\
%\bar{\textbf{S}}_{3+}(m) &= [\bar{\textbf{s}}_{3+}(n_{m})\:\: ... \:\: \bar{\textbf{s}}_{3+}(n_{m}+L-1)]^T, \\
%\tilde{\textbf{x}}_{3+}(m) &= [\bar{\boldsymbol{\alpha}}_{3+}(m-1)^T \bar{\textbf{S}}_{3+}(m)^T]^T , \\  
%\textbf{R}(m) &= [e^{j 2\pi \frac{3 f_{IF}}{f_s} (n_{m})} ... \:\: e^{j 2\pi \frac{3 f_{IF}}{f_s} (n_{m}+L-1)}]^T, \\
%\tilde{\textbf{x}}(m) &=  \textbf{x}(m) + \tilde{\textbf{x}}_{3+}(m)\circ \textbf{R}(m),
%\end{align}
%\normalsize
%where $(\circ)$ denotes element-wise multiplication.
While the above presentation describes the block-adaptive learning only at the IM3+ sub-band, extending the principle to higher-order IM sub-bands is straightforward, and thus not explicitly shown.
%{\color{red} ***22.10.2015 EDITED UP TO THIS POINT***}

\section{Implementation Aspects and Complexity Analysis}
\label{sec:Complexity}
One of the main advantages of the proposed decorrelation-based sub-band DPD technique is its reduced complexity compared to the classical full-band DPD processing, especially in scenarios where the CCs are widely spaced and thus very high speed ADCs and DACs are required in the classical solutions.
In this section, we first address some implementation aspects of the proposed sub-band DPD concept. We then provide a thorough comparison of the computing and hardware complexity perspectives between the full-band and the proposed sub-band DPDs. % We will use the term sub-band DPD to refer to the decorrelation-based sub-band DPD solution presented earlier. 
%However, before we start the complexity analysis we first present two possible implementation alternatives of the proposed sub-band DPD, followed by a comprehensive complexity analysis of the proposed DPD. 
Finally, some system power efficiency considerations are also presented, with particular focus on low-cost mobile devices.
%. The power budget with versus without using a DPD is demonstrated using practical digital signal processors available for mobile devices, thus demonstrating the practicality of the proposed DPD solution for small devices. 

\subsection{Sub-band DPD Implementation Alternatives}
Fig. \ref{fig:ImplementationAlternatives} shows two alternative architectures of the overall sub-band DPD processing with multiple IM sub-bands. 
%A single narrowband feedback receiver can be used, in which case the parameter learning for the multiple sub-band DPDs is done in a sequential manner, as explained earlier. This significantly reduces the overall complexity and cost of implementing the DPD. 
The first architecture shown in Fig. \ref{fig:Decorr_DPD_IM5_IM7_IM9} adds an upsampling and digital IF upconversion block after each sub-band DPD stage in order to digitally place the generated injection signals at proper intermediate frequencies. 
After adding the sub-band DPD outputs in the digital domain, a single wideband DAC per I and Q branch is used after which the signal is upconverted to RF and amplified by the PA. 

The second architecture in Fig. \ref{fig:Decorr_DPD_IM5_IM7_IM9_2} adds the outputs of the sub-band DPDs in the analog domain, implying that each sub-band DPD is followed by a narrow-band DAC per I and Q branch, together with an analog complex IF upconversion. A common I/Q RF modulator is then used for all sub-bands prior to the PA module. 

Both of these two architectures have their own advantages and disadvantages. In particular, if the carrier spacing between the component carriers is not very large,  architecture I is likely to be more suitable. On the other hand, when the carrier spacing starts to increase, using a single wideband DAC may not be efficient from cost and power consumption perspectives, so in this case architecture II is likely to be more attractive. However, some extra processing may be required in this architecture in order to achieve proper synchronization between the DACs. This forms an interesting topic for future research.

A common advantage of both architectures is that each sub-band DPD block can be switched on or off according to the prevailing emission levels and limits at the considered sub-bands. Such flexibility is not available in the classical full-band DPD solutions since, by design, a full-band predistorter always tries to linearize the full composite transmit band. With the sub-band DPD concept, the linearization can be flexibly tailored and optimized to those frequencies which are the most critical from the emission limits perspective.

%This flexibility can lead to clear power savings compared to the regular full-band DPD, as each sub-band DPD is consuming a certain amount of power much smaller than the full-band one. Thus, shutting down the DPD on those sub-bands which are not critical in terms of emission limits, allows us to optimize our allocated power according to the interference and spurious emission constraints that need to be met. Such flexibility is not available in the classical full-band DPD solutions since, by design, the predistorter always tries to linearize the full composite transmit band. The only tunable parameters in the full-band DPD are the predistorter nonlinearity order and memory depth. 

\subsection{Sub-band versus Full-band DPD Running Complexity}
In general, the computational complexity of any DPD can be classified into three main parts \cite{ComplexityAnalysis}: identification complexity, adaptation complexity and running complexity. The identification part is basically the estimation complexity of the DPD parameters, while the adaptation complexity includes the required processing in order to adapt to new operating conditions or device aging. 
%Both these two complexities are not very critical from a computational point of view since the former one can be done offline, while the latter occurs at a relatively low rate due to the slow speed of change of the PA operating characteristics compared to the data sampling rate. 
Finally, the running complexity, which is the most critical especially for mobile-type devices, involves the number of computations done per second while the DPD is operating. In this subsection, we will focus on the DPD running complexity in details, while the DPD parameter identification and feedback receiver complexity perspectives are shortly discussed in section \ref{sec:FB_complexity}.
For a quantitative comparison of the running complexities, we shall use the number of floating point operations (FLOPs) per sample, the number of DPD coefficients, and the required sample rate in the predistortion path as the main quantitative metrics. 

In general, the running complexity is divided into two main parts: the first is the basis function generation, and the second is the actual predistortion filtering using the basis functions \cite{ComplexityAnalysis}. 
The number of FLOPs required to perform these two operations is shown in Table \ref{tab:FLOPS} for ninth-order sub-band and full-band DPDs, respectively, where $N_1$ and $N_2$ are the corresponding memory depths per adopted basis function.
The full-band DPD architecture that we use also in our comparative performance simulations in Section V, and which is also widely applied otherwise, is based on the PH architecture with ninth order nonlinearity, while the sub-band DPD is based on architecture II shown in Fig. \ref{fig:Decorr_DPD_IM5_IM7_IM9_2}. 
%Moreover, in Table \ref{tab:FLOPS}, we have not considered the similarities between the basis functions of the different IM sub-bands that can lead to even more computational complexity reduction in actual sub-band DPD implementation. For example, generating the fifth order IM5+ basis function requires $4$ complex-complex multiplications, while if we can reuse the generated third order IM3+ basis function, we will only need $2$ complex-complex multiplications.
Frequency selectivity of the nonlinear PA is another important factor to be considered when comparing the two DPD architectures. This implies that when memory effects of the PA are considered, substantially longer filters are needed for the full-band DPD compared to the sub-band DPD for a certain performance requirement. In the complexity analysis, in the sub-band DPD case, we thus assume a memory depth of $1$ per basis function, while a memory depth of $3$ is assumed for the full-band PH DPD per basis function. 
%Adopting only a single tap per basis function in the sub-band DPD caseis shown to yield very good linearization performance with real PA measurements later in section \ref{sec:Simulations and Measurements}.

A substantial reduction in the needed sample rates can be achieved when adopting the sub-band DPD. As a concrete example, if we consider a challenging LTE-A UL scenario of two 1 MHz CCs with 20 MHz carrier spacing, the required sample rate by the full-band DPD is $9 \times (20+1) = 189 MSPS$, while for the sub-band DPD it becomes $9 \times 1 = 9 MSPS$.
\textcolor{black}{Consequently, as shown in Table \ref{tab:FLOPS}, a huge reduction in the number of FLOPs per second (FLOPS) can be achieved using the sub-band DPD}. Furthermore, the processing complexity of the sub-band DPD solution is clearly feasible for modern mobile device processing platforms, in terms of the GFLOPS count, while that of the classical fullband DPD is clearly infeasible. 
However, in some scenarios, when the carrier spacing between the CCs is decreased and/or the CC bandwidth increases, the benefit from using the sub-band DPD approach is reduced since the sample rates of the sub-band and full-band DPD will become more comparable. Notice also that, e.g., in interband CA cases, one can adopt a concurrent 2D-DPD \cite{S.A.BassamOct.2011,RoblinNov2013} to linearize the main carriers, which can then be easily complemented with the sub-band DPD processing to protect the own receiver if TX filtering does not offer sufficient isolation. Such a scenario could take place, e.g., in uplink Band 1 (1920-1980 MHz) + Band 3 (1710-1785 MHz) interband CA where it is technologically feasible to adopt a single multiband/multimode PA module for amplification. In such scenario, the IM3 sub-band is then directly at the own RX frequencies of Band 1 (2110-2170 MHz).

%Moreover, in many scenarios it is not required to suppress all the spurious intermodulation components at the same time. 
%For example, the transmit filter can already suppress some of those spurious intermodulations while some others may still be in-band. In other cases, spurious intermodulations at a particular sub-band could lead to own receiver desensitization, thus making the suppression of intermodulations at this particular sub-band critical while other sub-bands are not \cite{CommagCA_LTE,3GPP_duplexer}. 
%This flexibility is not existing in the full-band DPD, and can significantly reduce the complexity and power consumption of the sub-band DPD, which is critical especially for mobile devices and small-cell base-stations.
 
\subsection{Feedback Receiver and Parameter Estimation Complexity}
\label{sec:FB_complexity}
In addition to the complexity reduction in the DPD main processing path, the complexity of the feedback or observation receiver, used for DPD parameter estimation and adaptation, is also greatly reduced. In order to estimate, e.g., the parameters of the IM3+ sub-band DPD, we only need to observe the IM3+ sub-band at the PA output, instead of observing the full-band (including all the IM sub-bands) which is the case in the full-band DPD. 
This reduces the cost, complexity and power consumption in the feedback path allowing use of simpler instrumentation, in particular the ADC. Moreover, only a single observation receiver is required, even if linearizing multiple sub-bands, since the parameter learning and corresponding observation of the PA output can be done sequentially, in a per sub-band manner. 
Finally, in terms of the parameter estimation algorithmic complexity, the proposed decorrelation based solutions are extremely simple, when compared to any classical full-band DPD related method, such as least-squares based parameter fitting and indirect learning architecture \cite{GreenComm} that are commonly adopted. 

%{\color{red}In general, there can exist some impairments in the DPD feedback receiver due to the imperfections in the feedback analog components. For example, synchronization errors, observation RX I/Q imbalance, and observation RX DC offset can exist in the feedback receiver. The nonlinearity in the feedback path can be considered negligible, as the directional couple or the like providing the input to the observation receiver contains high intrinsic attenuation. Regarding the synchronization, I/Q imbalance, and DC offset impairments, they can be corrected at the receiver using well known algorithms in the literature. These impairments can exist in both full-band and sub-band DPD architectures alike.}

\subsection{Overall Transmitter Power Efficiency Perspectives}
When a DPD is adopted, a less linear but more efficient PA, that can operate near its saturation region, can generally be used. However, the overall power efficiency of the device is only improved if the extra power consumed by the DPD stage is less than the power savings due to increased PA efficiency. Here, we address this aspect from a mobile device perspective.

We consider a practical scenario where the transmit power at the output of the mobile PA is +26 dBm (i.e., 400 mW), stemming from 3GPP LTE-Advanced requirements \cite{3GPP,3GPP_CA_Emissions_1}, and assuming that the Tx duplexer filter and connector insertion losses are 3 dB. Then, good examples of practical PA power efficiency figures when operating in highly nonlinear or linear modes are around 35$\%$ and 20$\%$ (or even less), respectively. This means that the power consumed by the highly nonlinear PA is roughly 1150 mwatt, while the corresponding linear PA consumes roughly 2000 mwatt. In other words, adopting a highly power-efficient nonlinear PA saves 850 mwatt of power in this particular example.
Then, in order to suppress the spurious emissions at the PA output, in the nonlinear PA case, we adopt the sub-band DPD solution. 

%This sub-band DPD processing, in the example shown in Table \ref{tab:FLOPS}, requires approximately 1.78 GFLOPS and 1.55 GFLOPS for linearizing the IM3 and IM5 sub-bands respectively. 
In \cite{Qualcomm_v5_hw}, a state of the art 28 nm implementation of the Qualcomm Hexagon DSP capable of supporting up to 4.8 GFLOPS at 1.2 GHz is reported, which is more than enough for carrying out the needed sub-band DPD processing for the IM3 and IM5 sub-bands in our example. In the full-band DPD case, on the other hand, which requires 31.9 GFLOPS as shown in Table \ref{tab:FLOPS}, this is clearly insufficient. 
The power consumption of the DSP platform in \cite{Qualcomm_v5_hw} is shown to be approximately 100 mwatt, when running at 840 MHz (i.e., 3.36 GFLOPS assuming 4 FLOPS per cycle), which is again sufficient for linearizing the IM3 and IM5 sub-bands in our example. Thus, adopting the nonlinear PA already saves 850 milliwatt, when it comes to the PA interface, as explained above, while when complemented with the sub-band DPD processing for enhanced linearity, only 100 milliwatt of additional power is consumed. Thus, the overall power budget and power-efficiency are clearly in favor of using a highly non-linear PA, complemented with the sub-band DPD structure, even in a mobile device with non-contiguous uplink carrier aggregation. % scenarios like the one shown in Table \ref{tab:FLOPS}.
Furthermore, if the sub-band DPD processing is implemented using a dedicated hardware solution (e.g., a digital ASIC), an even more power-efficient DPD stage can most likely be realized. 
%{\color{red} *28.10.15 EDITED TILL HERE mv*}

\section{Simulation Results}
\label{sec:Simulations and Measurements}
In this section, a quantitative performance analysis of the proposed sub-band DPD solution is presented using Matlab simulations with practical models for mobile-like PAs designed for low-cost devices. 
% Practical RF measurements are also carried out using a commercial PA targeted particularly for mobile devices, showing the effectiveness of the proposed DPD solution in terms of spur mitigation in real-life scenarios.
%This section is organized as follows: first a quantitative comparison of the analytical DPD expressions presented in section \ref{sec:DPD_Analytic_Solutions} is performed, followed by a comparison of the fifth order decorrelation-based and fifth order inverse sub-band DPD solutions versus the third order counterparts presented in earlier works. The performance of the higher-order decorrelation-based sub-band DPD up to ninth order processing per IM sub-band is then demonstrated. Then, a quantitative comparison between the full-band and sub-band DPD is presented in terms of linearization performance and computational complexity using a practical noncontiguous transmission example with a real transmit filter. Finally, practical RF measurements are demonstrated using a commercial mobile PA, showing the effectiveness of the proposed DPD solution in terms of spur mitigation in real-life scenarios.
In general, we quantify the suppression of intermodulation power at the IM3 and IM5 sub-bands through the power ratios relative to the component carrier wanted signal power as shown in Fig. \ref{fig:PSD}, which are defined as
\begin{align}
IM3R &= 10 \log_{10} \frac{P_{wanted}}{P_{IM3}}, \:\:
IM5R &= 10 \log_{10} \frac{P_{wanted}}{P_{IM5}}. 
\end{align}
The inband transmit waveform purity is measured through the error vector magnitude (EVM), which is defined as
\begin{equation}
EVM_{\%} = \sqrt{\frac{P_{error}}{P_{ref}}} \times 100\%.
\end{equation}
Here, $P_{error}$ is the power of the error signal, while $P_{ref}$ is the reference power of the ideal symbol constellation. The error signal is defined as the difference between the ideal symbol values and the corresponding synchronized and equalized samples at the PA output, both normalized to have identical linear gains.
% In all EVM evaluations, linear distortion of the transmit chain is properly equalized prior to calculating the error signal.
%\cite{Dahlman4Gbook}.

\subsection{Comparison of Analytical IM3 Sub-band DPD Solutions}
\label{sec:Analytical_Sim}
In Section \ref{sec:DPD_Analytic_Solutions}, analytical reference expressions for the third-order inverse, minimum MSE, and decorrelation-based sub-band DPD solutions were presented. 
In this subsection, we shortly evaluate and compare the performance of these analytical solutions, assuming a third-order memoryless PA and known parameters, and focus only on the positive IM3 sub-band for simplicity. The memoryless PA model has been identified using a true mobile PA, with transmit power of +23dBm.
The main objective is to compare the performance of the three analytical solutions in terms of linearization performance, and thereon to verify that the decorrelation based solution is essentially identifcal to the minimum MSE solution. 

The signal used in this performance evaluation is composed of two 1MHz LTE-A UL SC-FDMA CCs with QPSK data modulation, and the CC spacing is 10MHz. The obtained results shown in
Table \ref{tab:IM3R_EVM_Analytical} demonstrate that the decorrelation-based solution is giving almost the same performance as the minimum MSE solution, and also that both of them are substantially better than the classical third-order inverse solution \cite{P.RoblinJan.2008} in terms of the IMD suppression at the considered sub-band. It can also be seen from Table \ref{tab:IM3R_EVM_Analytical} that the EVM is essentially not affected by the sub-band DPD processing. 
%In general, these results demonstrate that using the IM3 decorrelation-based sub-band DPD solution is essentially as good as the MMSE sub-band DPD solution for a third order PA from a linearization performance point of view. The advantage of using the decorrelation-based sub-band DPD over the MMSE solution is that it can be implemented using a simple hardware architecture as explained in section \ref{sec:DPD_adaptive_filter} and in \cite{ICASSP2014}. 

\begin{table}[t!]
\caption{Comparison of IM3+ suppression and EVM of third-order inverse, MMSE and decorrelation-based analytical sub-band DPD solutions at +23 dBm PA output power}
\centering
{\begin{tabular}{lll}\hline
                   & Positive IM3R [dBc]    & EVM [$\%$] \\\hline\hline
No DPD             & 28.8342   & 1.9711       \\\hline
Third-order Inverse      & 41.9230   & 1.8912       \\\hline
Analytical Decorrelation-based & 51.8248   & 1.8914       \\\hline
Analytical MMSE    & 51.8468   & 1.8917       \\\hline
\end{tabular}}{}
\label{tab:IM3R_EVM_Analytical}
\end{table}
%\begin{figure}[h!]
%\centering
%\vspace{0.5cm}
%\centerline{\includegraphics[width=0.5\textwidth]{./PSD_Analytical.eps}}
%\caption[]{Baseband equivalent of two 1 MHz LTE-A UL carriers with 12 MHz separation using a third order memoryless PA. MMSE and Decorrelating DPD solutions give almost the same performance which is better than the third order inverse solution.}
%\label{fig:PSD_Analytical}
%\end{figure}

\subsection{$P^{th}$-order Inverse vs. Decorrelation-based IM3 Sub-band DPD Solutions}
\label{sec:Inv_vs_decorr_sim}
% Extending the work done in \cite{ICASSP2014} to include higher order basis functions is one of the main contributions of this article as explained in details in section \ref{sec:DPD_adaptive_filter}. 
Next we evaluate and compare the performance of 5th-order inverse and decorrelation-based IM3 sub-band DPD solutions. The 5th-order inverse reference solution for IM3+ sub-band is derived in Appendix B. In these simulations, the transmit waveform is again composed of two 1 MHz LTE-A UL SC-FDMA component carriers with QPSK data modulation, and the CC spacing is 10 MHz. The PA model, in turn, is a memoryless 5th-order model whose parameters have been identified using a true mobile PA transmitting at +23 dBm. The 5th-order inverse DPD is using known parameters, while the decorrelation-based one is adopting the proposed sample-adaptive learning described in detail in Section IV.

\begin{figure}[t!]
\centering
%\vspace{0.5cm}
\centerline{\includegraphics[width=0.48\textwidth]{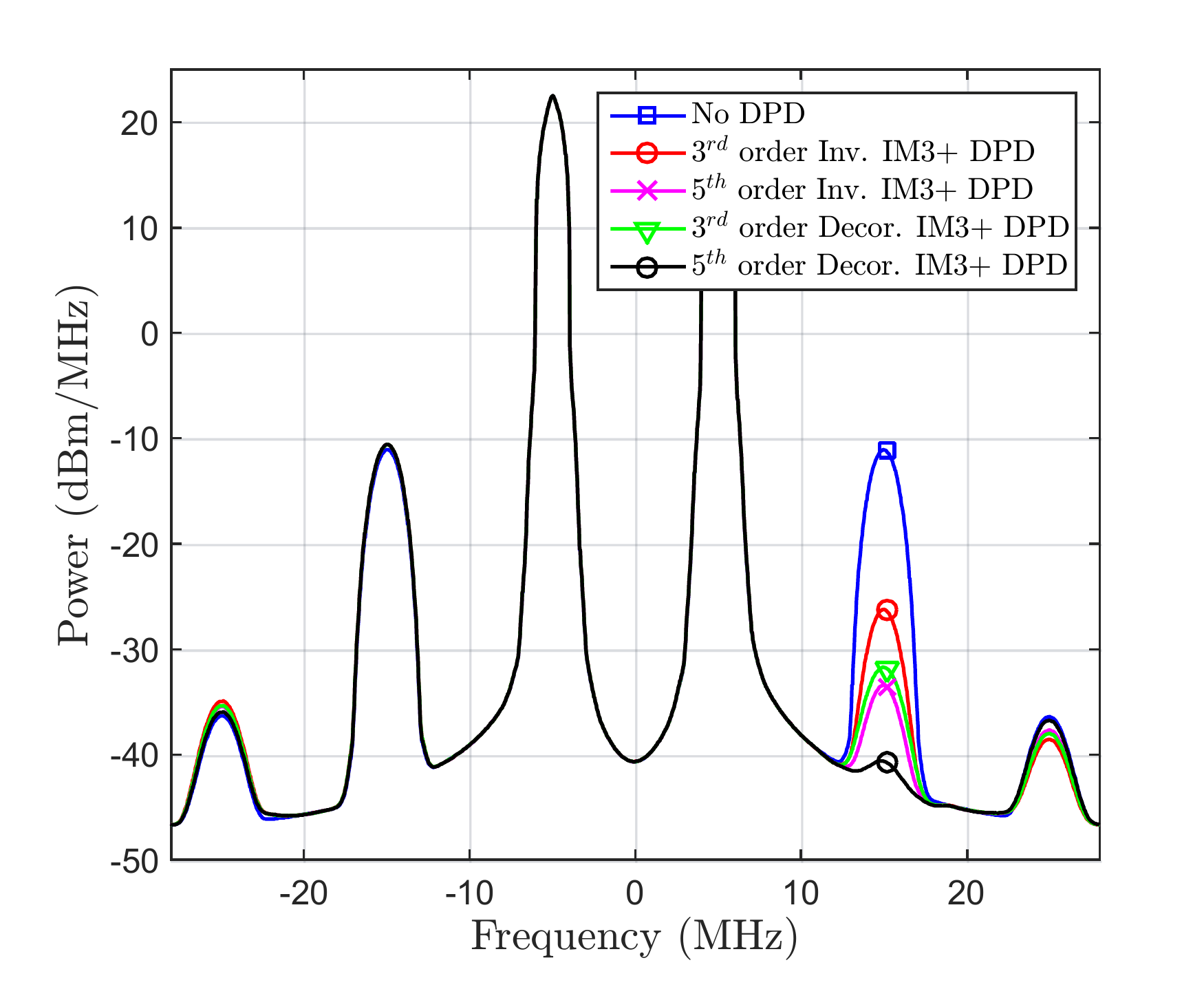}}
\caption[]{Baseband equivalent PA output spectrum at +23 dBm with two 1 MHz LTE-A UL carriers and 10 MHz CC separation using a fifth-order memoryless PA. $P^{th}$-order inverse and decorrelation-based IM3+ sub-band DPD solutions are compared, with both third-order and fifth-order nonlinear processing.}
\label{fig:PSD_Pth_Decorr}
\end{figure}

The PA output spectra with different solutions are illustrated in Fig. \ref{fig:PSD_Pth_Decorr}. It can clearly be seen that the decorrelation-based sub-band DPD substantially outperforms the $P^{th}$-order inverse based sub-band DPD, in both third and fifth-order cases, despite the fact that the $P^{th}$-order inverse solutions are using known parameters. The reason is that the inverse solutions cancel only the third and fifth order terms, as described in Appendix B, but do not suppress the other induced higher-order terms, which have structural similarity and correlation with the third and fifth-order basis functions. On the other hand, the proposed decorrelation-based solution takes this explicitly into account, and thus achieves clearly better spurious emission suppression. 
%In addition, no direct modeling for the PA is required in the decorrelation-based solution, while the fifth order inverse requires first to estimate the PA direct model before calculating the fifth order inverse sub-band DPD coefficients. 
The figure also illustrates that the performance of the fifth-order decorrelation-based sub-band DPD is clearly better than that of the third-order one.
 
In the remaining parts of this section, we shall focus on more detailed performance evaluations of the proposed decorrelation-based sub-band DPD solution, incorporating memory both in the PA and in the predistortion processing. From now on, we refer to the decorrelation-based sub-band DPD simply as 'sub-band DPD', to simplify the presentation.

\begin{figure}[t!]
\centering
%\vspace{0.5cm}
\centerline{\includegraphics[width=0.48\textwidth]{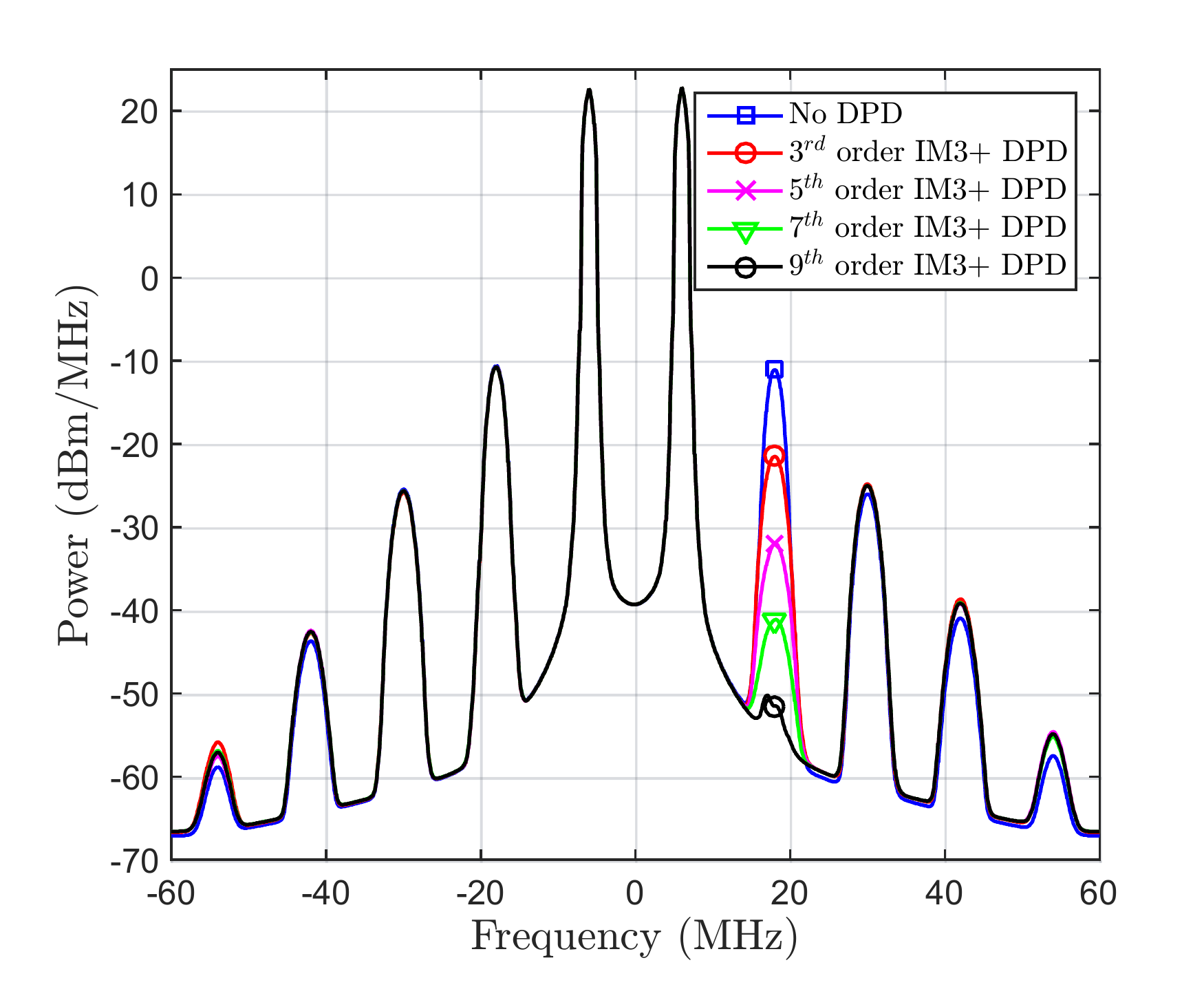}}
\caption[]{Baseband equivalent PA output spectrum at +24 dBm with two 1 MHz LTE-A UL carriers and 12 MHz CC separation using a ninth-order PA with memory. Different orders of the IM3+ sub-band DPD are compared, with memory depth equal to 1 per DPD SNL basis function.}
\label{fig:PSD_IM3}
\end{figure}

\begin{figure}[t!]
\centering
%\vspace{0.5cm}
\centerline{\includegraphics[width=0.48\textwidth]{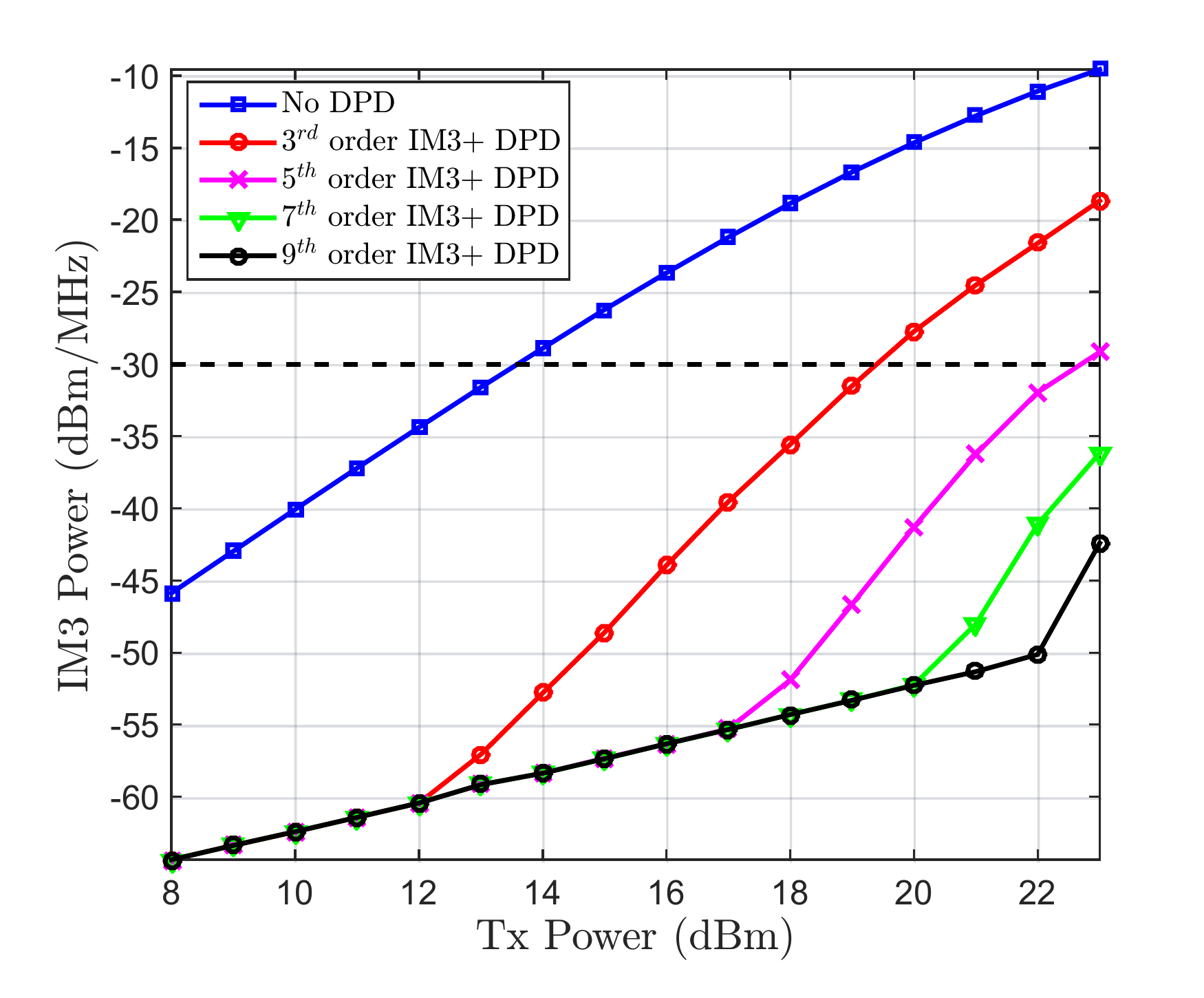}}
\caption[]{IM3 spurious emissions vs. TX power using different orders of the IM3+ sub-band DPD processing, with memory depth equal to 1 per DPD SNL basis function. A ninth-order PA with memory is used, and 2 dB Tx filter insertion loss is assumed.}
\label{fig:IM3_Pwr}
\end{figure}

\subsection{Performance of Proposed Higher-Order Sub-band DPD at IM3, IM5 and IM7}
In this subsection, we evaluate the performance of the proposed sub-band DPD in the more realistic case of having memory in the PA. The PA model is a 9th-order parallel Hammerstein model, with four memory taps per branch, and the parameters have been identified using measurements with a true mobile PA transmitting at +24 dBm. The sub-band DPD structure contains now also memory, with two taps ($N=1$) per basis function. Block-adaptive learning principle is adopted, with 200 blocks each containing 1000 samples. The transmit waveform is otherwise identical to earlier cases, but the CC separation is now 12 MHz.

\begin{table*}
\caption{Comparison of quantitative running complexity and linearization performance of full-band versus sub-band DPD. Two 1 MHz LTE-A UL carriers with QPSK data modulation and 20 MHz carrier spacing are used.}
\centering
{\begin{tabular}{llll|llll}\hline
				& \multicolumn{3}{c}{\textbf{DPD Running Complexity}} \vline &  \multicolumn{3}{c}{\textbf{Transmitter Performance}}    \\\hline
                          &  Coeffs  & Fs [MSPS] & GFLOPS   & EVM [\%]  & Positive IM3R [dBc]  & Positive IM5R [dBc]  & Output Power [dBm]   \\\hline
\textbf{No DPD}           &  N/A		 & N/A       & N/A      & 1.2527    & 35.5744              & 56.4272 							& +20   \\\hline
\textbf{Full-Band ILA DPD}&  20 	   & 189       & 31.941   & 0.1058    & 61.9428              & 63.9842              & +19   \\\hline
\rowcolor{Gray}
\textbf{IM3+ Sub-Band DPD}&  8       & 9         & 0.891    & 1.2489    & 68.3291              & N/A      				    & +20   \\\hline
\rowcolor{Gray}
\textbf{IM5+ Sub-Band DPD}&  6       & 9         & 0.774    & 1.2529    & N/A                  & 71.9823  					  & +20   \\\hline
\end{tabular}}{}
\label{tab:IM3_IM5_EVM}
\end{table*}

%In this paper higher order sub-band DPD solutions with orders higher than five have also been developed. 
Fig. \ref{fig:PSD_IM3} shows the effectiveness of the proposed higher-order sub-band DPDs, here processing the IM3+ sub-band, compared to the basic third-order solution presented earlier in \cite{CROWNCOM2014}. Up to 40 dB suppression for the IM3 spurious emissions is shown when adopting ninth-order processing, something that has not been reported before in any prior works using a practical PA model with memory effects. Fig. \ref{fig:IM3_Pwr} shows then how the IM3 spurious emission level changes with varying the TX power and using different sub-band DPD orders. The IM3 spurious emissions are clearly below the general spurious emission limit (-30 dBm/MHz) even at very high TX powers up to +23 dBm, when using a seventh-order, or higher, sub-band DPD.

\begin{figure}[t!]
\centering
%\vspace{0.5cm}
\centerline{\includegraphics[width=0.48\textwidth]{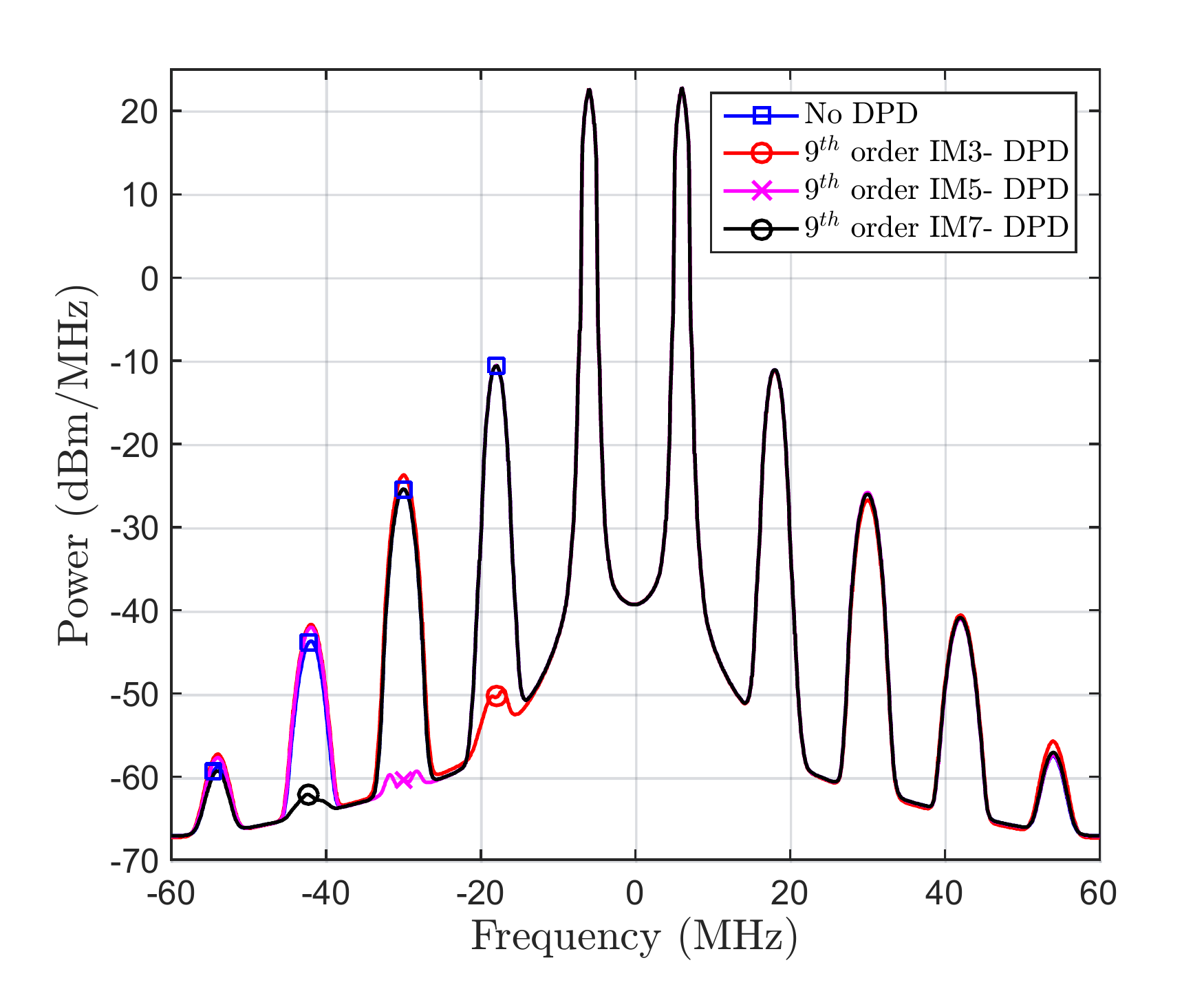}}
\caption[]{Baseband equivalent of two 1 MHz LTE-A UL carriers with 12 MHz separation using a ninth order PH PA extracted from a real mobile PA at +24 dBm. Ninth order negative IM3, IM5, and IM7 sub-band DPD solutions are shown, with memory depth equal to 1 per DPD SNL basis function.}
\label{fig:PSD_IM3_5_7_9}
\end{figure}

Another main contribution in this paper is the extension of the IM3 sub-band DPD solution to include also higher-order IM sub-bands. Fig. \ref{fig:PSD_IM3_5_7_9} shows the performance of the negative IM5 and IM7 sub-band DPDs, in addition to the negative IM3 sub-band DPD, when ninth-order nonlinear processing is adopted in the sub-band DPDs. Up to 40 dB, 35 dB, and 18 dB of suppression is achieved at the negative IM3, IM5, and IM7 sub-bands, respectively. This shows the effectiveness of the proposed solutions in processing and suppressing also the spurious emissions at higher IM sub-bands. 
%This proves the flexibility and effectiveness of the solution in mitigating spurs at different IM sub-bands with different nonlinearity orders considered. 
%Thus, allowing the designers of mobile devices and small-cell base-stations to optimize the sub-band DPD complexity, power consumption, and cost according to the linearization requirements. 

\begin{figure*}[t!]
\centering
\subfigure[]{
\includegraphics[width=0.7\textwidth]{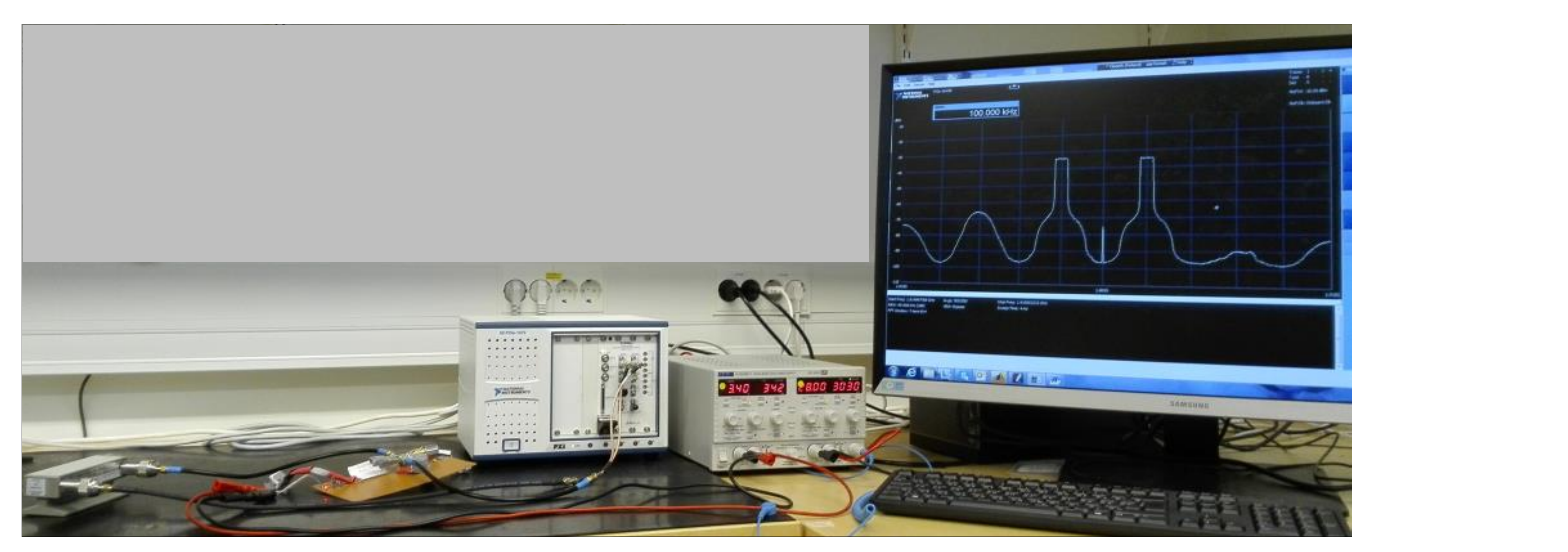}
\label{fig:MeasurementPhoto}
}
\hfill
\subfigure[]{
\includegraphics[width=0.7\textwidth]{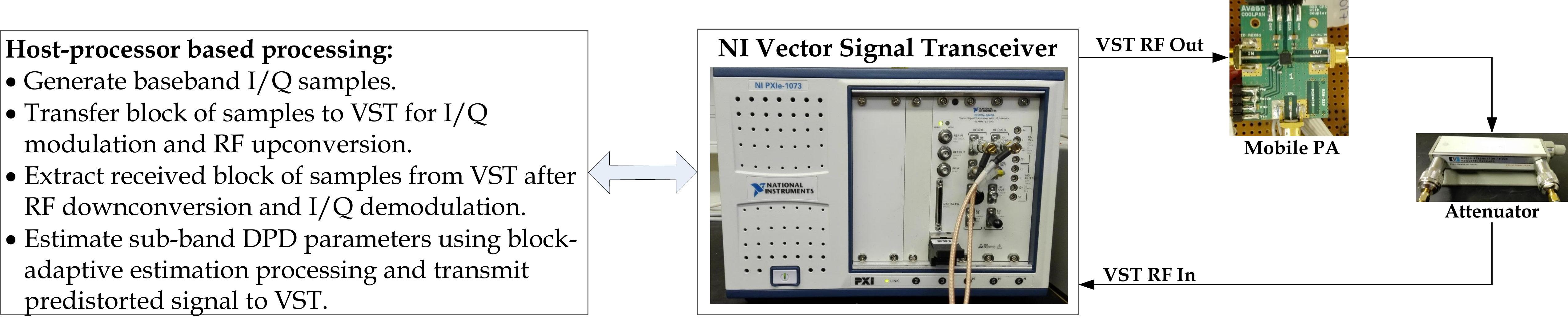}
\label{fig:MeasurementSetup}
}
\caption{Hardware setup used in the RF measurements for testing and evaluating the proposed sub-band DPD.}
\label{fig:MeasPhoto}
\end{figure*}

\subsection{Full-Band versus Sub-band DPD Complexity and Performance Analysis}

In this subsection, we compare the performance of our proposed sub-band DPD technique against that of a classical full-band DPD adopting parallel Hammerstein based wideband linearization and indirect learning architecture (ILA). The evaluation considers the linearization performance, complexity, and transmitter efficiency utilizing our earlier complexity analysis results reported in Section \ref{sec:Complexity}. The ninth-order full-band PH ILA DPD uses 100k samples for parameter learning, per ILA iteration, with a total of 2 ILA iterations, and the number of memory taps per PH branch is 4. 
The ninth-order block-adpative sub-band DPD, in turn, uses also a total of 200k samples with block size $M = 1000$, and adopts two memory taps per DPD SNL basis function.

The results are collected in Table \ref{tab:IM3_IM5_EVM}, which shows that in addition to the significantly lower complexity measured by the number of GFLOPS, the sub-band DPD achieves better linearization performance in terms of spurious IMD suppression at the considered IM3 and IM5 sub-bands.
On the other hand, the full-band DPD outperforms the sub-band DPD in terms of the inband distortion mitigation (i.e., EVM). 
This is expected, since the full-band DPD linearizes the whole transmit band, including the main CCs and the IMD spurious emissions. However, the EVM with the sub-band DPD is only around $1.25\%$ which is by far sufficient for modulations at least up to $64$-QAM. \textcolor{black}{Additionally, the full-band DPD based on the ILA structure requires an additional 1 to 2 dB back off to guarantee stable operation, which is not required in the sub-band DPD}. Thus the transmitter becomes more power efficient when using the sub-band DPD as shown in Table \ref{tab:IM3_IM5_EVM}. For fairness, it is to be acknowledged that a full-band DPD can typically enhance the EVM and ACLR, while the sub-band DPD concept is specifically targeting only the spurious emissions.

%A practical transmit filter that suppresses the negative IM terms is also used in our simulations leaving only the positive IM terms in-band, which are in turn suppressed by the DPD.
 
%Multiple sub-band DPDs can be used, and as explained in section \ref{sec:Complexity}, their combined complexity is in many scenarios still significantly lower than the full-band DPD since not all sub-band DPDs need to operate in parallel. 
%For example in the scenario shown in Fig. \ref{fig:PSD_Full_Sub_Band_DPD}, the transmit filter suppresses the negative IMD terms leaving only the positive ones in-band. In this example, the IMD at the positive IM3 sub-band is violating the spurious emission limit at -30 dB/1 MHz, thus it is the most critical from a transmitter perspective. 
%However, for demonstration, the positive IM5 sub-band DPD is also shown in order to compare the suppression of the spurs in the IM5 sub-band when using sub-band versus full-band DPD. 

\section{RF Measurement Results}
\label{sec:RF Measurement Results}
%\begin{figure}[t!]
%\centering
%\vspace{0.5cm}
%\centerline{\includegraphics[width=0.47\textwidth]{./MeasurementSetup.pdf}}
%\caption[]{RF measurement setup using National Instruments Vector Signal Transceiver and a practical mobile PA. Matlab is used for baseband digital signal processing.}
%\label{fig:MeasurementSetup}
%\end{figure}
In order to further demonstrate the operation of the proposed sub-band DPD solution, we next report results of comprehensive RF measurements using a commercial LTE-Advanced mobile terminal PA together with a vector signal transceiver (VST), which is implementing the RF modulation and demodulation. The actual sub-band DPD processing and parameter learning algorithms are running on a host processor.
 
The ACPM-5002-TR1 mobile power amplifier used in our measurements is designed for LTE-A UL band 25 (1850-1915~MHz), with 29 dB gain. The National Instruments (NI) PXIe-5645R VST includes both a vector signal generator (VSG), and a vector signal analyzer (VSA) with 80 MHz instantaneous bandwidth. In these experiments, the digital baseband waveform is divided into 50 blocks of size $M = 10k$ samples which are first generated locally on the host processor, and then transferred to the VSG to perform RF I/Q modulation at the desired power level at the PA input.
The VST RF output is then connected to the input port of the external power amplifier, whose output port is connected to the VST RF input through a 40 dB attenuator, implementing the observation receiver as illustrated also in Fig. \ref{fig:MeasPhoto}. 
The VSA performs RF I/Q demodulation to bring the signal back to baseband. The baseband I/Q observation block is then filtered to select the IM3 sub-band which is  used for block-based sub-band DPD learning, after proper alignment with the locally generated basis functions, as explained in Section \ref{sec:block_based_DPD}. The sub-band DPD block size $M$ used in these experiments is $10k$, and the DPD memory depth $N$ is $1$.

In general, two different intra-band LTE-A CA RF measurement examples are demonstrated in this subsection. The first experiment demonstrates the violation of the spurious emission limit due to the inband emission of the IM3+ spur, thus not being attenuated by the TX filter. The second experiment demonstrates an own RX desensitization example, in FDD transceiver context, where the IM3+ spur is located at the own RX band (1930-1995~MHz) and is not sufficiently attenuated by the duplexer TX filter. Notice that in principle, when tackling specifically the own RX desensitization problem with the proposed sub-band DPD solution, the main receiver of the FDD device could potentially be used as the observation receiver (i.e., learning the sub-band DPD coefficients without an extra auxiliary/observation receiver). However, such an approach would indeed be applicable only in the own RX desensitization case, while when mitigating other harmful emissions, an auxiliary observation receiver would anyway have to be adopted. Thus, in these RF measurements, we adopt the auxiliary receiver based approach, while consider further developments of the main receiver based parameter learning as an important topic for our future work.

\begin{figure}[t!]
\centering
%\vspace{0.5cm}
\centerline{\includegraphics[width=0.48\textwidth]{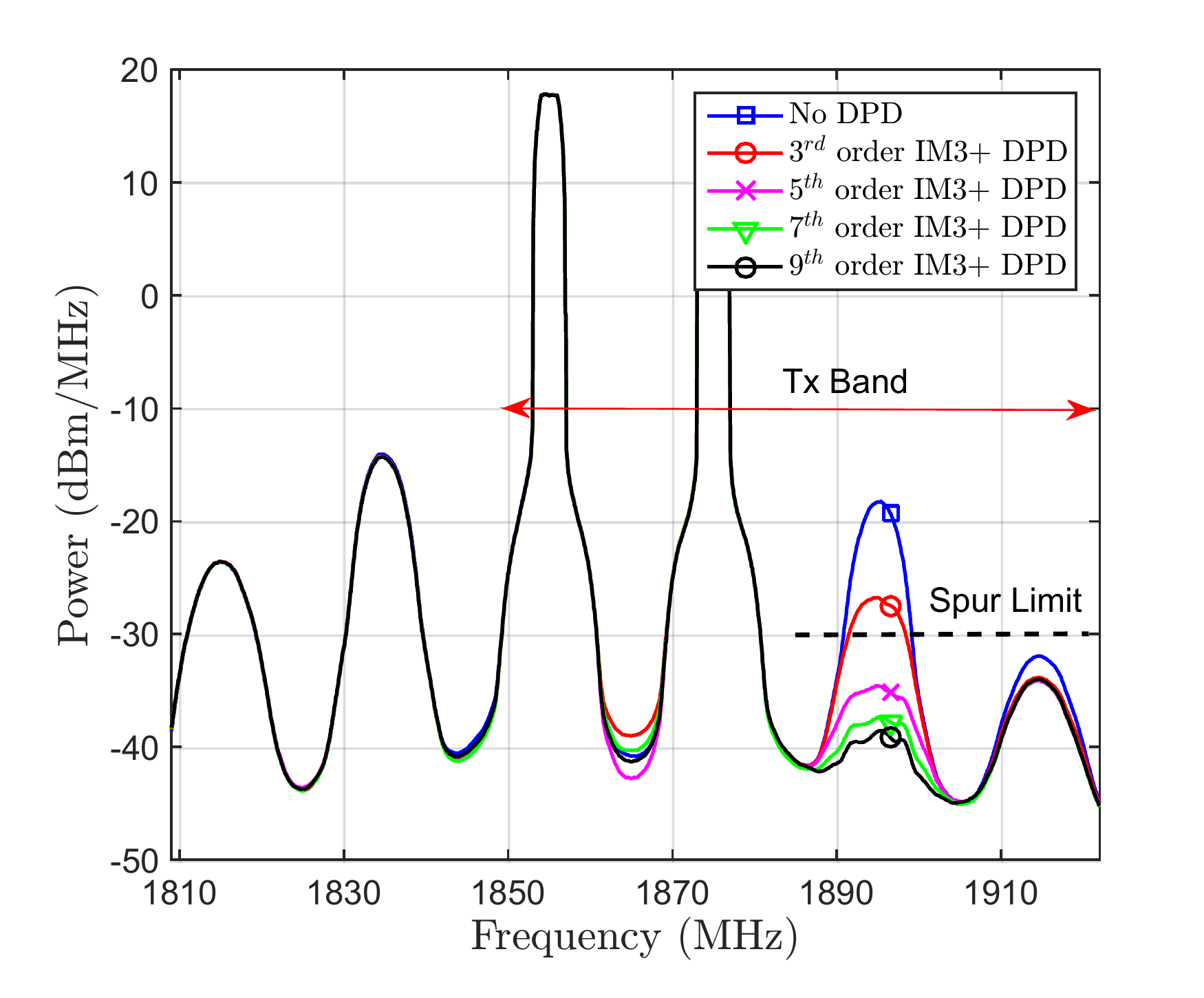}}
\caption[]{An LTE-A band 25 RF measurement example at the UE PA output showing the gain from using an IM3+ sub-band DPD. IM3 spur reduction with third, fifth, seventh, and ninth-order sub-band DPDs are demonstrated, using a real commercial mobile PA operating at +25 dBm. An LTE-A UL CA signal with two 3 MHz CCs and 20 MHz carrier spacing is used.}
\label{fig:PSD_Lab_IM3}
\end{figure}

\begin{figure}[t!]
\centering
%\vspace{0.5cm}
\centerline{\includegraphics[width=0.48\textwidth]{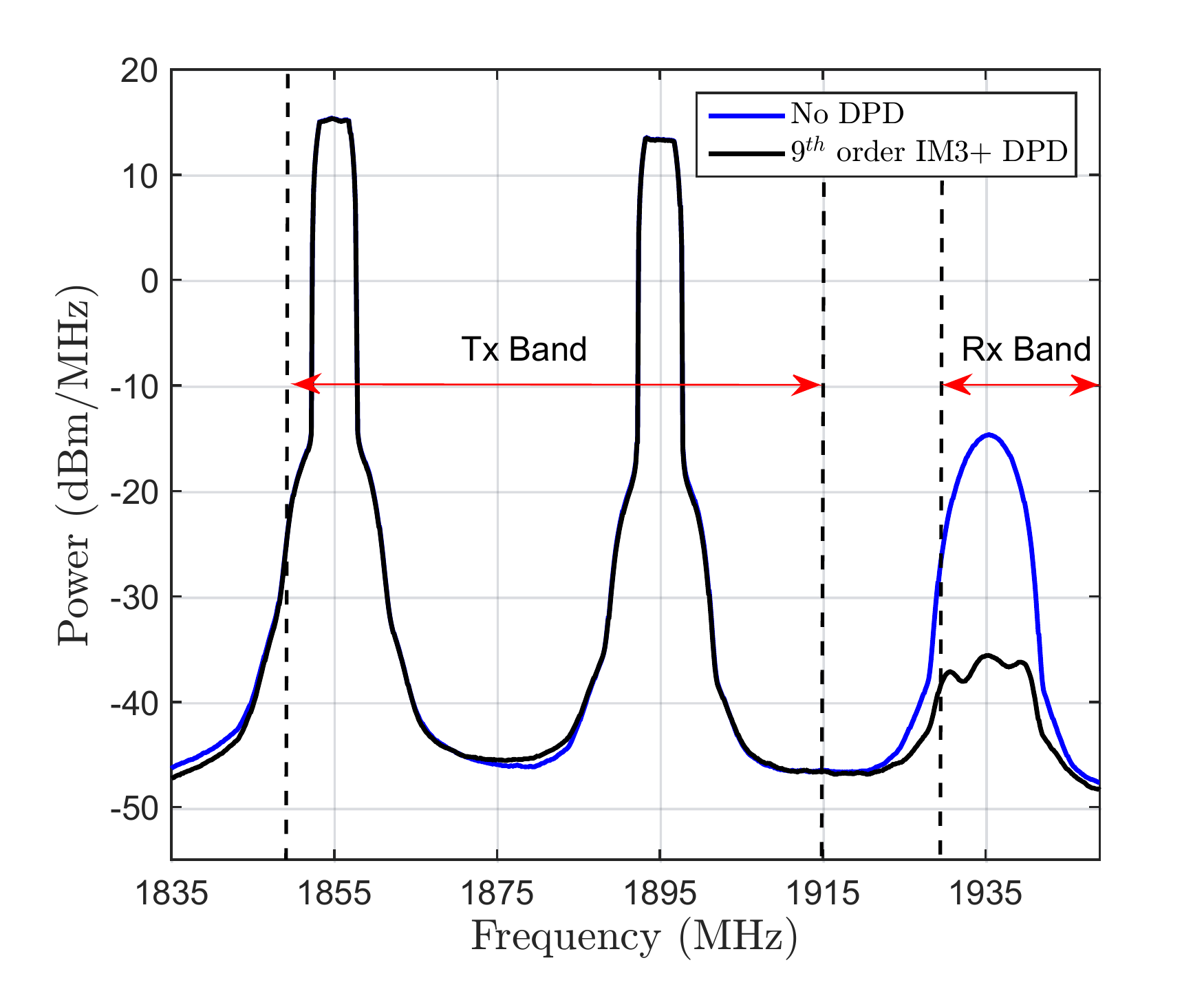}}
\caption[]{An LTE-A band 25 RF measurement example at the UE PA output showing the gain from using a ninth-order IM3+ sub-band DPD when the IM3+ is falling at own RX band. An LTE-A UL CA signal with two 5 MHz CCs and 40 MHz carrier spacing is used with a real commercial mobile PA operating at +25 dBm.}
\label{fig:PSD_RXSENS_LAB}
\end{figure}

\subsection{Spurious Emission Limit Violation} 
Fig. \ref{fig:PSD_Lab_IM3} shows the measured power spectral density at the PA output using the IM3 sub-band DPD with different orders. The adopted waveform is an intra-band CA LTE-A UL signal with 3 MHz per CC and 20 MHz carrier spacing. 
The power level at the PA output in this example is +25 dBm.
The intermodulation distortion at the IM3+ sub-band is emitted inband, since the total TX band covers 1850-1915~MHz, and is clearly violating the spurious emission limit (-30 dBm/MHz) when no DPD is used. When using the sub-band DPD, the spurious emission level is well below the emission limit, given that at least 5th-order processing is deployed. \textcolor{black}{In general, more than 20 dB of measured spurious emission suppression is achieved with the ninth order IM3 sub-band DPD, thus giving more than 10 dB additional gain compared to the basic third-order sub-band DPD as shown in Fig. \ref{fig:PSD_Lab_IM3}}. Notice that there is no need for predistorting the IM3- sub-band in this example, since those emissions will be filtered out by the TX filter.

\subsection{Own Receiver Desensitization} 
Fig. \ref{fig:PSD_RXSENS_LAB} illustrates another LTE-A intra-band CA UL band 25 example, now with 5 MHz CC bandwidths and 40 MHz carrier spacing. \textcolor{black}{The duplexing distance at LTE-A band 25 is only 80 MHz, thus the IM3+ spur at 1935 MHz in this example is falling on the own RX band, and therefore potentially desensitizing the own receiver.} % of the first CC which is at 1855 MHz.
The power level at the PA output in this example is +25 dBm.

\begin{figure}[t!]
\centering
%\vspace{0.5cm}
\centerline{\includegraphics[width=0.48\textwidth]{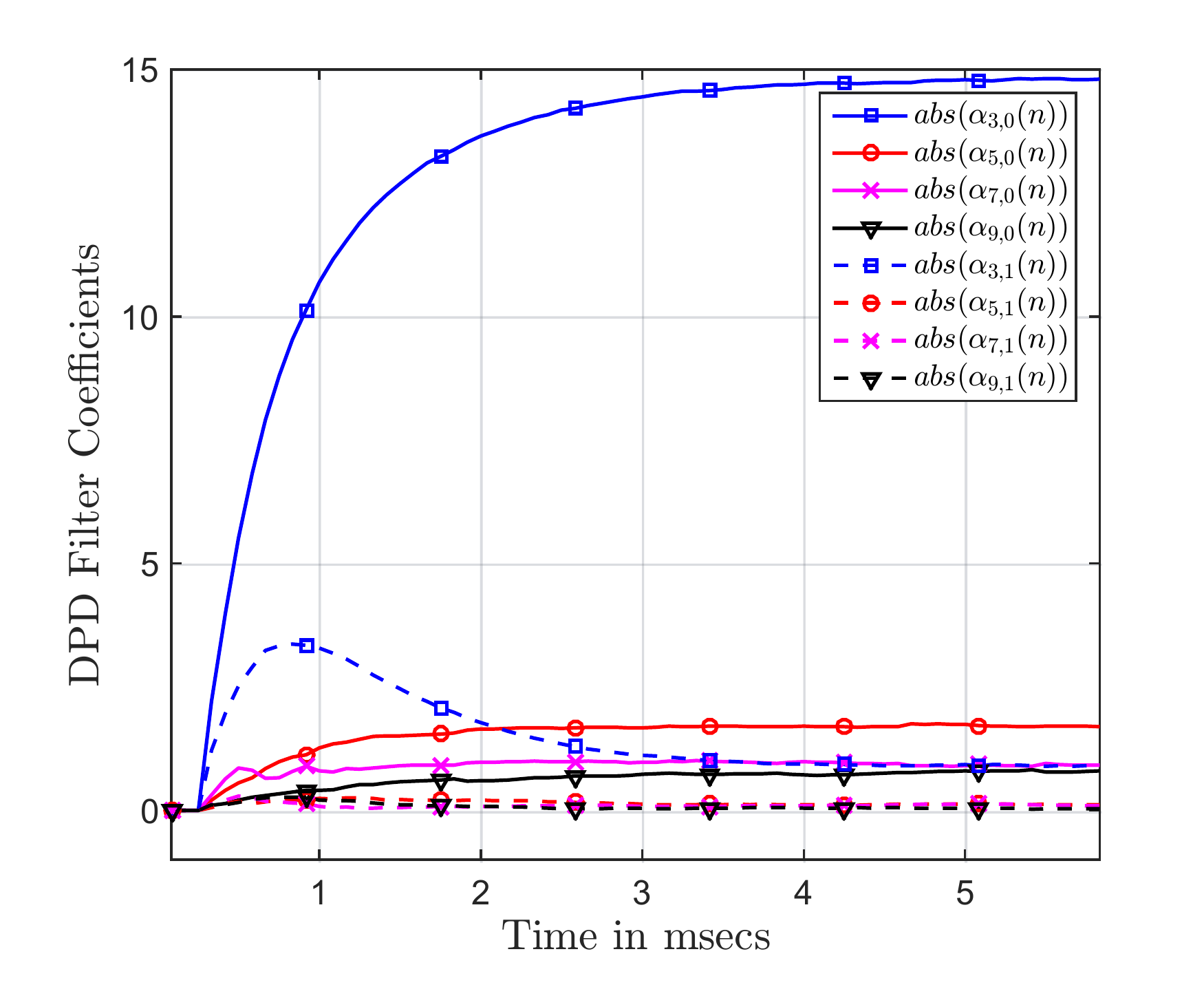}}
\caption[]{Convergence of the ninth-order sub-band DPD coefficients, with memory depth $N = 1$ per basis function, when the positive IM3 sub-band is considered and block-adaptive learning is deployed. An RF measurement example with two 5 MHz LTE-A UL CCs and 40 MHz carrier spacing is used with a real commercial mobile PA operating at +25 dBm.}
\label{fig:DPD_Coeff_9th}
\end{figure}

Then, by adopting a ninth-order IM3+ sub-band DPD, more than 20 dB of spurious emission suppression is achieved, as can be seen in Fig. \ref{fig:PSD_RXSENS_LAB}. Again, two taps ($N=1$) per basis function are used, and block-adaptive parameter learning is deployed. Fig. \ref{fig:DPD_Coeff_9th} shows the real-time convergence of the sub-band DPD coefficients for the third, fifth, seventh, and ninth-order basis functions, respectively. It can be seen that the coefficients converge in a stable manner in a real RF environment due to the orthogonalization of the SNL basis functions, as explained in Section \ref{sec:DPD_adaptive_filter}.

\begin{figure}[t!]
\centering
%\vspace{0.5cm}
\centerline{\includegraphics[width=0.48\textwidth]{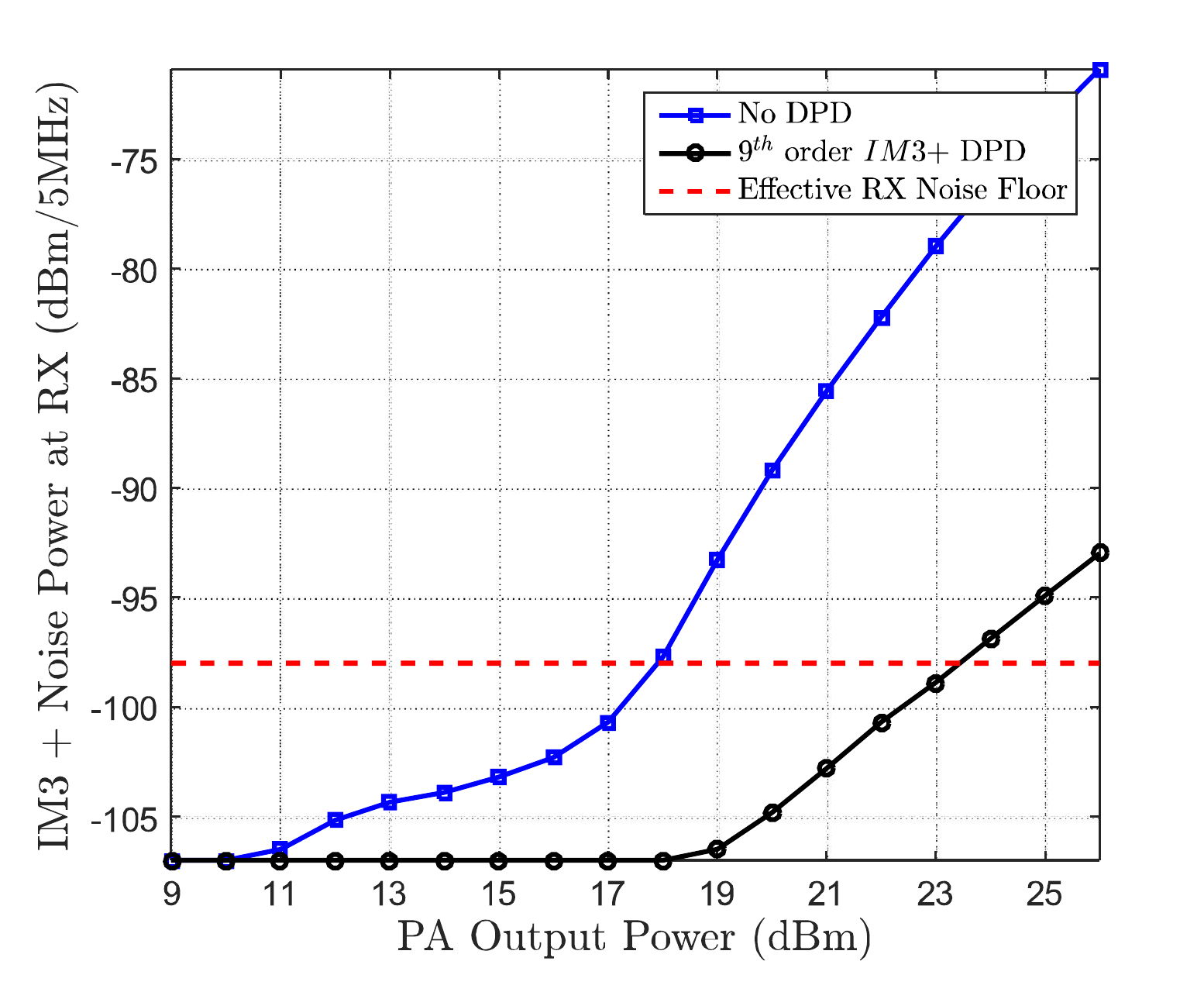}}
\caption[]{Integrated IM3+ sub-band power at own RX LNA input over 5 MHz vs. PA output power using ninth order IM3+ sub-band DPD processing, with memory depth equal to 1 per DPD SNL basis function. A commercial mobile PA operating at LTE band 25 is used in the RF measurements, and 65 dB duplexer TX filter attenuation \cite{Avago_Duplexer} is assumed at the RX band.}
\label{fig:IM3_PWR_RxBand}
\end{figure}

\textcolor{black}{In general, assuming a UE duplexer TX filter with 65 dB attenuation at the own RX band (e.g., ACMD-6125 Band 25 LTE-A UE Duplexer), the integrated power of the IM3+ spur at the RX band without DPD will be approximately -73 dBm/5MHz}. This is 25 dB above the effective RX noise floor when assuming 9 dB UE RX noise figure (NF) \cite{3GPP2}.
This would thus cause significant own receiver desensitization that could lead to a complete blocking of the desired RX signal. On the other hand, when the proposed IM3+ sub-band DPD is deployed, the integrated power of the IM3+ spur at the  RX band will be approximately -95 dBm/5MHz, which is only 3 dB above the effective RX noise floor, as shown also in Fig. \ref{fig:IM3_PWR_RxBand}. 
Though the residual spur is still slightly above the effective RX noise floor, the sensitivity degradation is substantially relaxed, despite operating at a maximum PA output power of +25 dBm. 
We elaborate on this further in Fig. \ref{fig:IM3_PWR_RxBand}, showing the integrated power of the IM3+ spur at the RX LNA input while changing the PA output power level. With a ninth order sub-band DPD, we can transmit up to +18 dBm PA output power with an effectively perfectly linear TX, in terms of IM3+ spur level, compared to +10 dBm PA output power without DPD. Additionally, with DPD, the integrated IM3+ power is less than the effective RX noise floor up to +23 dBm PA output power, while without DPD it is already 20 dB above the noise floor at the same power level.
%only 1 dB back off is required to bring the IM3 interference below the reference sensitivity level, compared to 7 dB back off when no DPD is used. This significantly enhances the power efficiency and UL coverage in LTE-A FDD transceivers.

\section{Conclusions}
In this article, a novel low-complexity sub-band digital predistortion (DPD) solution was proposed for suppressing unwanted spurious emissions in non-contiguous spectrum access. Novel decorrelation based adaptive parameter learning methods were also formulated, allowing efficient estimation and tracking with low computational complexity. All algorithm derivations and modeling were carried out in the general case of having memory in the PA as well as in the sub-band DPD processing. Different nonlinear distortion and processing orders, beyond classical third-order cases, were also reported. The proposed technique can find application in suppressing inband spurs which would violate the spurious emission limit, suppressing out-of-band spurs falling, e.g., on the own receiver band, or in protecting primary user transmissions in cognitive radio systems.
%The proposed DPD was mainly targeting the spurious intermodualtion distortions between component carriers in noncontiguous transmission schemes. Higher order nonlinearities and higher order intermodualtion bands were considered.
A quantitative complexity analysis was presented, comparing the proposed solution to conventional full-band DPD solutions available in the literature. The performance was evaluated in a comprehensive manner, showing excellent linearization performance despite the considerably reduced complexity compared to classical full-band solutions. Finally, extensive RF measurement results using a commercial LTE-Advanced mobile power amplifier were reported, evidencing up to 22 dB suppression of the most problematic third-order spurious emissions. 

\section*{Appendix A: Analytical MMSE Solution for Third-Order IM3 Sub-band DPD}
Here, we derive the analytical minimum mean-square error solution shown in (32), which is used as a reference solution in simulations in Section \ref{sec:Analytical_Sim}. We first define the so-called error signal as $e(n)=\tilde{y}_{IM3_+}(n)$, since with ideal predistortion the IM3+ sub-band signal would be zero, and thus the optimization means minimizing the power of this error signal. 
From (\ref{eq:ThirdOrderAnalysis}), the error signal reads
%The error signal $e(n)$ which we want to minimize its power $\tilde{y}_{IM3_+}(n)$ is the PA output at the IM3+ sub-band with the DPD included, which reads 

\small
\begin{align}
e(n) &= (f_3 + f_1 \alpha) u(n) + 2 f_3 \alpha (|x_1(n)|^2 + |x_2(n)|^2) u(n) \nonumber\\
&+ f_3 |\alpha|^2 \alpha |x_1(n)|^4 |x_2(n)|^2 u(n),
\end{align}
\normalsize
where $f_1$ and $f_3$ are the third-order memoryless PA model parameters, $\alpha$ denotes the DPD coefficient to be optimized, and $x_1(n)$ and $x_2(n)$ are the baseband equivalents of the two component carriers. 
The statistical expectation $\E[|e(n)|^2]=\E[e(n)e^*(n)]$, assuming that the component carrier signals $x_1(n)$ and $x_2(n)$ are statistically independent, and ignoring some high-order vanishingly small terms, then reads 

\small
\begin{align}
&\E[e(n)e^*(n)] = \E_{42} \left[|f_3|^2 + |f_1|^2|\alpha|^2 + \alpha f_1 f_3^* + \alpha^* f_1^* f_3 \right] \nonumber\\ 
&+ (2 \E_{62} + 2 \E_{44}) \left[|f_3|^2 (\alpha + \alpha^*) + |\alpha|^2(f_1 f_3^* + f_1^* f_3) \right] \nonumber\\
&+ (4 \E_{46} + 8 \E_{64} + 4 \E_{82}) |f_3|^2 |\alpha|^2, 
\label{eq:MeanSquaredError}
\end{align}
\normalsize
where $\E_{ij}$ is used as a shorthand notation for $\E[|x_1|^i] \E[|x_2|^j]$.
Now, differentiating (\ref{eq:MeanSquaredError}) with respect to $\alpha$, yields 

\small
\begin{align}
\frac{\partial}{\partial\alpha}\E[e(n)e^*(n)] &= \E_{42}\left[\alpha^* |f_1|^2 + f_1 f_3^* \right] \nonumber\\
&+ (2 \E_{62} + 2 \E_{44}) \left[|f_3|^2 + \alpha^* (f_1 f_3^* + f_1^* f_3) \right] \nonumber\\
&+ (4 \E_{46} + 8 \E_{64} + 4 \E_{82}) |f_3|^2 \alpha^*.
\label{eq:MeanSquaredErrorDiff}
\end{align}
\normalsize
Then, setting (\ref{eq:MeanSquaredErrorDiff}) to zero and solving for $\alpha$ yields the optimal MMSE DPD parameter, given by

\small 
\begin{align}
&\alpha_{MMSE} = \nonumber\\
&\dfrac{-\left[f_1 f_3^* \E_{42} + 2 |f_3|^2 (\E_{62} + \E_{44})\right]^*}{\left[|f_1|^2 \E_{42} + 4 \mathbb{R}(f_1 f_3^*)(\E_{62} + \E_{44}) 
+ 4 |f_3|^2 (\E_{46} + 2 \E_{64} + \E_{82})\right]^*},
\end{align}
\normalsize
where $\mathbb{R}(.)$ denotes the real-part operator. This concludes the proof.
%First we evaluate the expression for $e(n)e^*(n)$, while removing the sample index $n$ in $x_1(n)$ and $x_2(n)$ for the sake of presentation compactness as
%\small
%\begin{align}
%& e(n)e^*(n) = |x_1|^4 |x_2|^2 \left[|f_3|^2 + |f_1|^2|\alpha|^2 + \alpha f_1 f_3^* + \alpha^* f_1^* f_3 \right] \nonumber\\ 
%&+ (2|x_1|^6 |x_2|^2 + 2|x_1|^4 |x_2|^4) \left[|f_3|^2 (\alpha + \alpha^*) + |\alpha|^2(f_1 f_3^* + f_1^* f_3) \right] \nonumber\\
%&+ (4|x_1|^4 |x_2|^6 + 8|x_1|^6 |x_2|^4 + 4|x_1|^8 |x_2|^2) |f_3|^2 |\alpha|^2 \nonumber\\
%&+ (4|x_1|^{10} |x_2|^4 + 4|x_1|^8 |x_2|^6) |f_3|^2 |\alpha|^4 \nonumber\\
%&+ |x_1|^8 |x_2|^4\left[\alpha |f_3|^2 |\alpha|^2 + \alpha^* |f_3|^2 |\alpha|^2 + |\alpha|^4 (f_1 f_3^* + f_1^* f_3)\right] \nonumber\\
%&+ |x_1|^{12} |x_2|^6 |f_3|^2 |\alpha|^6 
%\label{eq:SquaredError}
%\end{align}
%\normalsize

%Then, we can directly operate with the statistical expectation operator $\E[.]$ to (\ref{eq:SquaredError}). Assuming that the component carrier signals $x_1$ and $x_2$ are statistically independent, we obtain 
%The expectation of the last $3$ rows in (\ref{eq:SquaredError}) have been neglected as they are vanishingly small compared to other terms. 

\section*{Appendix B: Analytical Fifth-Order Inverse Based IM3 Sub-band DPD Solution}
Here, we derive the $5^{th}$ order inverse solution for the IM3+ sub-band DPD, to be used as a reference solution in the simulations in Section \ref{sec:Inv_vs_decorr_sim}. The output of a memoryless $5^{th}$ order polynomial PA model is

\small
\begin{align}
y(n) &= f_1 x(n) + f_3 |x(n)|^2 x(n) + f_5 |x(n)|^4 x(n), \label{eq:PA_5th_NoMem}
\end{align}
\normalsize
where $f_1$, $f_3$ and $f_5$ are the polynomial coefficients, and $x(n)$ is the composite baseband equivalent input signal, as given in (\ref{eq:PA_In}). Through direct substitution of (\ref{eq:PA_In}) in (\ref{eq:PA_5th_NoMem}), the baseband equivalent IM3+ distortion term, located at three times the IF frequency, can be extracted and reads

\small
\begin{align}
y_{IM3_+}(n) &= f_3 (x_2^*(n)x_1^2(n)) \nonumber\\
&+ 2 f_5 |x_1(n)|^2 (x_2^*(n)x_1^2(n)) \nonumber\\
&+ 3 f_5 |x_2(n)|^2 (x_2^*(n)x_1^2(n)). \label{eq:IM3_out_5th_NoMem} 
\end{align}
\normalsize
Stemming from the signal structure in (\ref{eq:IM3_out_5th_NoMem}), the sub-band DPD injection signal is composed of three basis functions of the form $x_2^*(n)x_1^2(n)$, $|x_1(n)|^2 x_2^*(n)x_1^2(n)$ and $|x_2(n)|^2 x_2^*(n)x_1^2(n)$. In case of fifth-order inverse based sub-band DPD, these basis functions are multiplied by proper coefficients such that all the distortion terms at the IM3+ sub-band at the PA output, up to order five, are cancelled. Thus, incorporating such DPD processing, yet with arbitrary coefficients, the composite baseband equivalent PA input signal reads

\small
\begin{align}
\begin{split}
\tilde{x}(n) &= x_1(n) e^{j 2\pi \frac{f_{IF}}{f_s} n} + x_2(n) e^{-j 2\pi \frac{f_{IF}}{f_s} n} \\
&+ \alpha_{3,inv} (x_2^*(n)x_1^2(n)) e^{j 2\pi \frac{3 f_{IF}}{f_s} n} \\
&+ \alpha_{51,inv} |x_1(n)|^2 (x_2^*(n)x_1^2(n)) e^{j 2\pi \frac{3 f_{IF}}{f_s} n} \\
&+ \alpha_{52,inv} |x_2(n)|^2 (x_2^*(n)x_1^2(n)) e^{j 2\pi \frac{3 f_{IF}}{f_s} n}, \label{eq:PA_In_FifthDPD}
\end{split}
\end{align}
\normalsize
where the subscript $inv$ in the coefficients is emphasizing the $P^{th}$-order inverse based solution.
Substituting (\ref{eq:PA_In_FifthDPD}) in (\ref{eq:PA_5th_NoMem}), and extracting the third and fifth order IM3+ terms, yields

\small
\begin{align}
\begin{split}
\tilde{y}&_{IM3_+}(n) = (f_1 \alpha_{3,inv} + f_3) x_2^*(n)x_1^2(n) \\
&+ (f_1 \alpha_{51,inv} + 2f_3 \alpha_{3,inv} + 2f_5) |x_1(n)|^2 (x_2^*(n)x_1^2(n)) \\
&+ (f_1 \alpha_{52,inv} + 2f_3 \alpha_{3,inv} + 3f_5) |x_2(n)|^2 (x_2^*(n)x_1^2(n)). \label{eq:IM3_terms}
\end{split}
\end{align}
\normalsize
From (\ref{eq:IM3_terms}), we now can easily obtain the fifth-order inverse coefficients that null the third and fifth-order terms, yielding

%\small
%\begin{align}
%\begin{split}
%\alpha_{3,inv}  = -\frac{f_3}{f_1} , \:\:\: 
%\alpha_{51,inv} = 2 \frac{f_3^2}{f_1^2} - 2 \frac{f_5}{f_1}, \:\:\: 
%\alpha_{52,inv} = 2 \frac{f_3^2}{f_1^2} - 3 \frac{f_5}{f_1}. \label{eq:Fifth_Order_Inverse_Sol}
%\end{split}
%\end{align}
%\normalsize
\small
\begin{align}
\begin{split}
\alpha_{3,inv}  &= -\frac{f_3}{f_1} , \nonumber\\
\alpha_{51,inv} = 2 \frac{f_3^2}{f_1^2} - 2 \frac{f_5}{f_1}, \:&\:
%\nonumber\\
\alpha_{52,inv} = 2 \frac{f_3^2}{f_1^2} - 3 \frac{f_5}{f_1}. \label{eq:Fifth_Order_Inverse_Sol}
\end{split}
\end{align}
\normalsize
This concludes the derivation. %This $5^{th}$-order inverse solution is used as a reference solution in some of the performance evaluations in Section \ref{sec:Simulations and Measurements}.
% show that the $5^{th}$ order decorrelation-based sub-band DPD solution outperforms the $5^{th}$ order inverse one, a conclusion that has earlier been shown for $3^{rd}$ order sub-band DPD in \cite{ICASSP2014}.

%\color{red}{
%\bibliographystyle{IEEEbib}
%\bibliographystyle{IEEEtran}
%\bibliography{Ref}

\begin{thebibliography}{10}

\bibitem{LTE_A_Parkvall}
S.~Parkvall, A.~Furuskar, and E.~Dahlman,
\newblock ``{Evolution of LTE toward IMT-advanced},''
\newblock {\em IEEE Commun. Mag.}, vol. 49, no. 2, pp. 84--91, February 2011.

\bibitem{3GPP}
{\em {LTE {E}volved {U}niversal {T}errestrial {R}adio {A}ccess {(E-UTRA)}
  {U}ser {E}quipment {(UE)} radio transmission and reception, 3GPP TS 36.101
  V12.4.0 (Release 12)}}, June 2014.

\bibitem{TransmitterArchitectureCA}
S.A. Bassam, W.~Chen, M.~Helaoui, and F.M. Ghannouchi,
\newblock ``Transmitter architecture for {CA}: {C}arrier {A}ggregation in
  {LTE}-{A}dvanced systems,''
\newblock {\em IEEE Microw. Mag.}, vol. 14, no. 5, pp. 78--86, July 2013.

\bibitem{CommagCA_LTE}
C.~Park, L.~Sundstr{\"o}m, A.~Wallen, and A.~Khayrallah,
\newblock ``Carrier aggregation for {LTE}-advanced: Design challenges of
  terminals,''
\newblock {\em IEEE Commun. Mag.}, vol. 51, no. 12, pp. 76--84, Dec. 2013.

\bibitem{3GPP_CA_Emissions_1}
R4-121205,
\newblock ``{Way forward for non-contiguous intraband transmitter aspects.''},
%\newblock 3GPP Tech. {R}ep., Nokia, Feb. 2013.
\newblock 3GPP Tech. {R}ep., Feb. 2013.

\bibitem{3GPP_CA_Emissions_2}
R4-124353,
\newblock ``{Non-contiguous intraband unwanted emission.''},
%\newblock 3GPP Tech. {R}ep., Nokia, Feb. 2013.
\newblock 3GPP Tech. {R}ep., Feb. 2013.
% \bibitem{LaehteensuoMay2013}
% T.~L{\"a}hteensuo,
% \newblock ``{Linearity Requirements in {LTE A}dvanced Mobile Transmitter.},''
% \newblock M.S. thesis, Tampere University of Technology, Tampere, Finland., May
%   2013.
%\textcolor{red}{
\bibitem{TI_DPD}
H.~Gandhi, D.~Greenstreet, and J.~Quintal,
\newblock ``{Digital Radio Front-End strategies provide game-changing benefits
  for small cell base stations},''
\newblock Tech. {R}ep., Texas Instruments, May 2013.
\bibitem{GreenComm}
L.~Guan and A.~Zhu,
\newblock ``Green communications: Digital predistortion for wideband {RF} power
  amplifiers,''
\newblock {\em IEEE Microw. Mag.}, vol. 15, no. 7, pp. 84--89, Dec. 2014.
\bibitem{3GPP_duplexer}
% Ericsson and ST-Ericsson, 
R4-123797,
\newblock ``{UE reference sensitivity requirements with two UL carriers.},''
% \newblock 3GPP Tech. {R}ep., Ericsson and ST-Ericsson, Feb. 2013.
\newblock 3GPP Tech. {R}ep., Feb. 2013.

\bibitem{ITU}
{\em International Telecommunication Union Radio Communication Sector,
  Recommendation {ITU-R} {SM}.329-12 Unwanted emissions in the spurious domain.}.
\bibitem{RX_Desens_Cancellation_Chao}
Chao Yu, Wenhui Cao, Yan Guo, and A.~Zhu,
\newblock ``{Digital Compensation for Transmitter Leakage in Non-Contiguous
  Carrier Aggregation Applications With FPGA Implementation},''
\newblock {\em IEEE Trans. Microw. Theory Techn.}, vol. 63, no. 12, pp.
  4306--4318, Dec 2015.
\bibitem{RX_Desens_Cancellation_Adnan}
A.~Kiayani, M.~Abdelaziz, L.~Anttila, V.~Lehtinen, and M.~Valkama,
\newblock ``{Digital Mitigation of Transmitter-Induced Receiver Desensitization
  in Carrier Aggregation FDD Transceivers},''
\newblock {\em IEEE Trans. Microw. Theory Techn.}, vol. 63, no. 11, pp.
  3608--3623, Nov 2015.
\bibitem{S.A.BassamOct.2011}
S.A. Bassam, F.M. Ghannouchi, and M.~Helaoui,
\newblock ``2-{D} {D}igital {P}redistortion (2-{D}-{DPD}) architecture for
  concurrent dual-band transmitters,''
\newblock {\em IEEE Trans. Microw. Theory Techn.}, vol. 59, no. 10, pp. 2547--2553, Oct.
  2011.
\bibitem{RoblinNov2013}
P.~Roblin, C.~Quindroit, N.~Naraharisetti, S.~Gheitanchi, and M.~Fitton,
\newblock ``Concurrent linearization,''
\newblock {\em IEEE Microw. Mag.}, vol. 14, no. 7, pp. 75--91, Nov. 2013.
\bibitem{SingleFB_DPD}
Y.~Liu, J.~Yan, and P.~Asbeck,
\newblock ``Concurrent dual-band digital predistortion with a single feedback
  loop,''
\newblock {\em IEEE Trans. Microw. Theory Techn.}, vol. 63, no. 5, pp.
  1556--1568, May 2015.
\bibitem{P.RoblinJan.2008}
P.~Roblin, S.~K. Myoung, D.~Chaillot, Y.~G. Kim, A.~Fathimulla, J.~Strahler,
  and S.~Bibyk,
\newblock ``Frequency-selective predistortion linearization of {RF} power
  amplifiers,''
\newblock {\em IEEE Trans. Microw. Theory Techn.}, vol. 56, no. 1, pp. 65--76, Jan.
  2008.
\bibitem{J.KimJan.2013}
J.~Kim, P.~Roblin, D.~Chaillot, and Z.~Xie,
\newblock ``A generalized architecture for the frequency-selective digital
  predistortion linearization technique,''
\newblock {\em IEEE Trans. Microw. Theory Techn.}, vol. 61, no. 1, pp. 596--605, Jan.
  2013.
\bibitem{S.A.BassamAug.2012}
S.A. Bassam, M.~Helaoui, and F.M. Ghannouchi,
\newblock ``Channel-selective multi-cell digital predistorter for multi-carrier
  transmitters,''
\newblock {\em IEEE Trans. Microw. Theory Techn.}, vol. 60, no. 8, pp. 2344--2352, Aug.
  2012.
\bibitem{ICASSP2014}
M.~Abdelaziz, L.~Anttila, A.~Mohammadi, F.~Ghannouchi, and M.~Valkama,
\newblock ``Reduced-complexity power amplifier linearization for carrier
  aggregation mobile transceivers,''
\newblock in {\em IEEE International Conference on Acoustics, Speech, and
  Signal Processing}, pp. 3908--3912, May 2014.
\bibitem{CROWNCOM2014}
M.~Abdelaziz, L.~Anttila, J.R. Cavallaro, S.~Bhattacharyya, A.~Mohammadi, F.M.
  Ghannouchi, M.~Juntti, and M.~Valkama,
\newblock ``Low-complexity digital predistortion for reducing power amplifier
  spurious emissions in spectrally-agile flexible radio,''
\newblock in {\em 9th International Conference on Cognitive Radio Oriented
  Wireless Networks}, pp. 323--328, June 2014.
\textcolor{black}{
\bibitem{ASILOMAR_ABDELAZIZ}
M.~Abdelaziz, C.~Tarver, K.~Li, L.~Anttila, R.~Martinez, M.~Valkama, J.~R.~Cavallaro,
\newblock ``{Sub-band digital predistortion for noncontiguous transmissions: Algorithm development and real-time prototype implementation},''
\newblock in {\em 49th Asilomar Conference on Signals, Systems and Computers}, pp. 1180--1186, Nov. 2015.
\bibitem{COMMAG_ABDELAZIZ}
M.~Abdelaziz, Z.~Fu, L.~Anttila, A.~M.~Wyglinski and M.~Valkama,
\newblock ``Digital predistortion for mitigating spurious emissions in spectrally agile radios,''
\newblock {\em IEEE Commun. Mag.}, vol. 54, no. 3, pp. 60--69, March 2016.
\bibitem{TCOM_ABDELAZIZ}
Z.~Fu, L.~Anttila, M.~Abdelaziz, M.~Valkama, and A.~M.~Wyglinski,
\newblock ``{Frequency-Selective Digital Predistortion for Unwanted Emission Reduction},''
\newblock {\em  IEEE Trans. Commun.}, vol. 63, no. 1, pp. 254--267, Jan. 2015.}
\bibitem{ComplexityAnalysis}
A.S. Tehrani, H.~Cao, S.~Afsardoost, T.~Eriksson, M.~Isaksson, and C.~Fager,
\newblock ``A comparative analysis of the complexity/accuracy tradeoff in power
  amplifier behavioral models,''
\newblock {\em IEEE Trans. Microw. Theory Techn.}, vol. 58, no. 6, pp. 1510--1520, June
  2010.

\bibitem{OrthPloyDPD}
H.~Qian, S.~Yao, H.~Huang, and W.~Feng,
\newblock ``A low-complexity digital predistortion algorithm for power
  ampliﬁer linearization,''
\newblock {\em IEEE Trans. Broadcast.}, vol. 60, no. 4, pp. 670--678, Dec.
  2014.

\bibitem{Iterative_Orth}
W.~Hoffmann,
\newblock ``Iterative algorithms for {Gram} {Schmidt} orthogonalization,''
\newblock {\em Computing}, vol. 41, no. 4, pp. 335--348, 1989.

\bibitem{Qualcomm_v5_hw}
M.~Saint-Laurent et~al.,
\newblock ``A 28 nm {DSP} powered by an on-chip {LDO} for high-performance and
  energy-efficient mobile applications,''
\newblock {\em IEEE J. Solid-State Circuits}, vol. 50, no. 1, pp. 81--91, Jan
  2015.

% \bibitem{Avago_Duplexer}
% {\em Avago Band 25 LTE-Advanced UE Duplexer, ACMD-6125. Available at:
%   http://docs.avagotech.com/docs/AV02-4214EN}.

\bibitem{3GPP2}
{\em {LTE {E}volved {U}niversal {T}errestrial {R}adio {A}ccess {(E-UTRA)} {RF
  system scenarios}, 3GPP TR 36.942 V10.2.0 (Release 10)}}, May 2011.

\end{thebibliography}
%}

\begin{IEEEbiography}[{\includegraphics[width=1in,height=1.25in,clip,keepaspectratio]{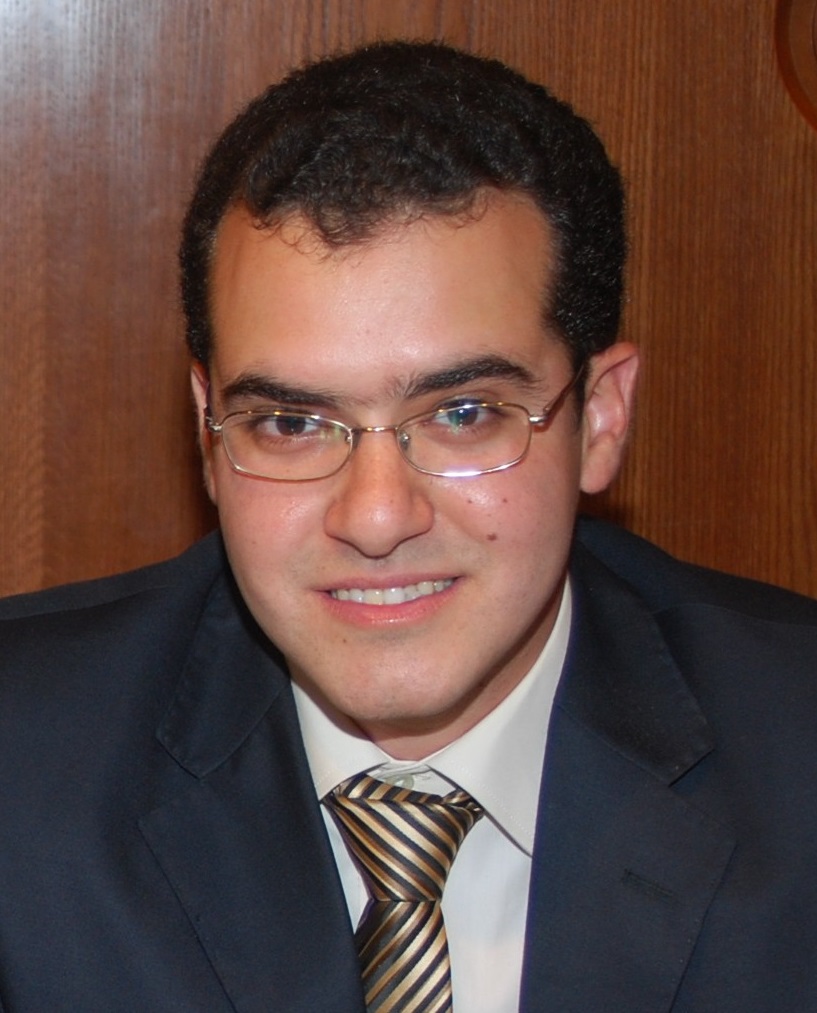}}]{Mahmoud Abdelaziz}
(S'13) received the B.Sc. degree (with honors) and M.Sc. degree in Electronics and Electrical Communications Engineering from Cairo University, Egypt, in 2006 and 2011. He is currently pursuing the Doctoral degree at Tampere University of Technology, Finland where he works as a researcher with the Department of Electronics and Communications. From 2007 to 2012 he has been working as a communication systems and signal processing engineer at Newport Media, as well as other companies in the wireless industry. His research interests include statistical and adaptive signal processing in flexible radio transceivers, wideband digital pre-distortion, and cognitive radio systems.
\end{IEEEbiography}
\begin{IEEEbiography}[{\includegraphics[width=1in,height=1.25in,clip,keepaspectratio]{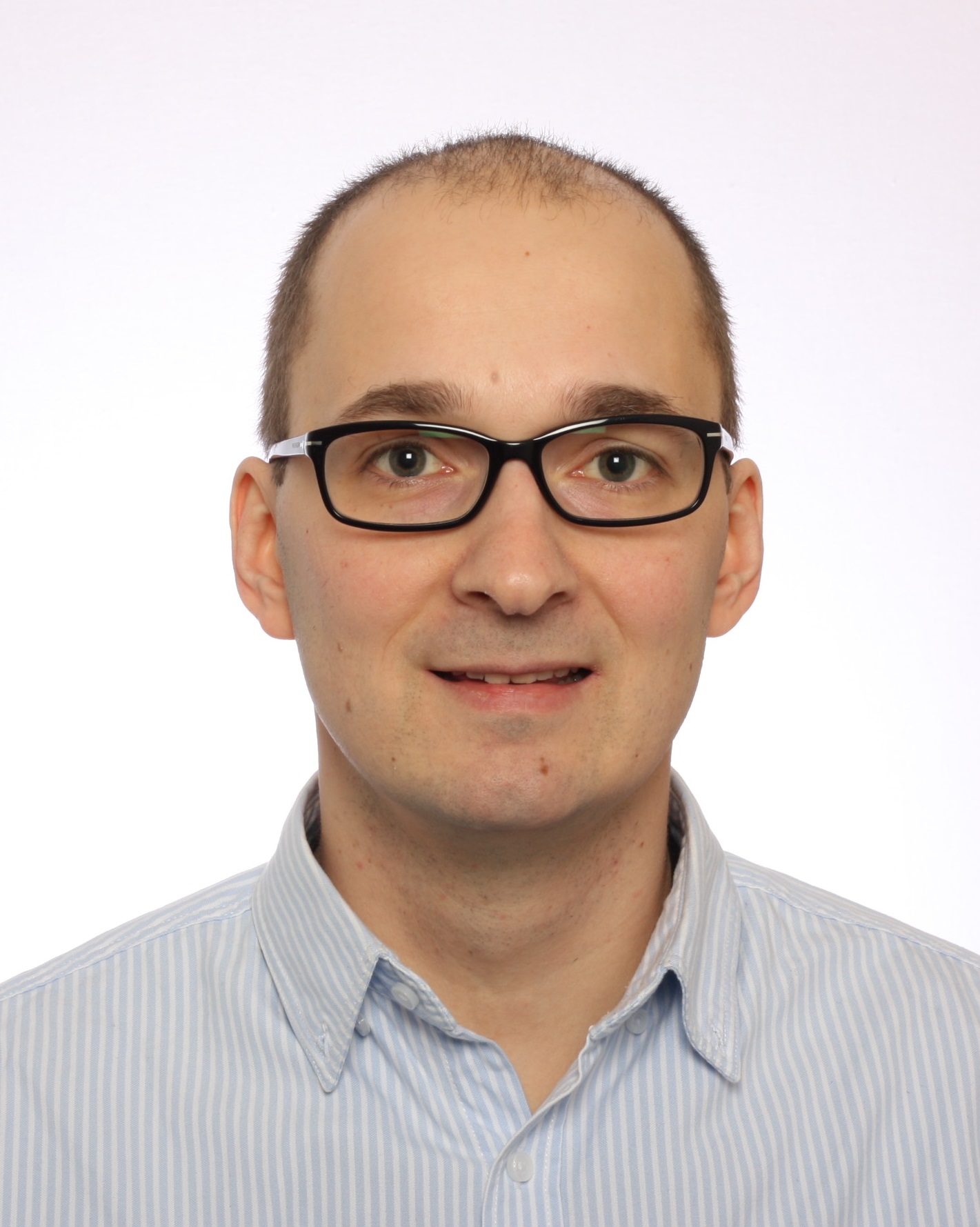}}]{Lauri Anttila}
(S'06, M'11) received the M.Sc. degree and D.Sc. (Tech) degree (with honors) in electrical engineering from Tampere University of Technology (TUT), Tampere, Finland, in 2004 and 2011. Currently, he is a Senior Research Fellow at the Department of Electronics and Communications Engineering at TUT. His research interests are in signal processing for wireless communications, in particular radio implementation challenges in 5G cellular radio and full-duplex radio, flexible duplexing techniques, and transmitter and receiver linearization. He has co-authored over 60 peer reviewed articles in these areas, as well as two book chapters.
\end{IEEEbiography}
\begin{IEEEbiography}[{\includegraphics[width=1in,height=1.25in,clip,keepaspectratio]{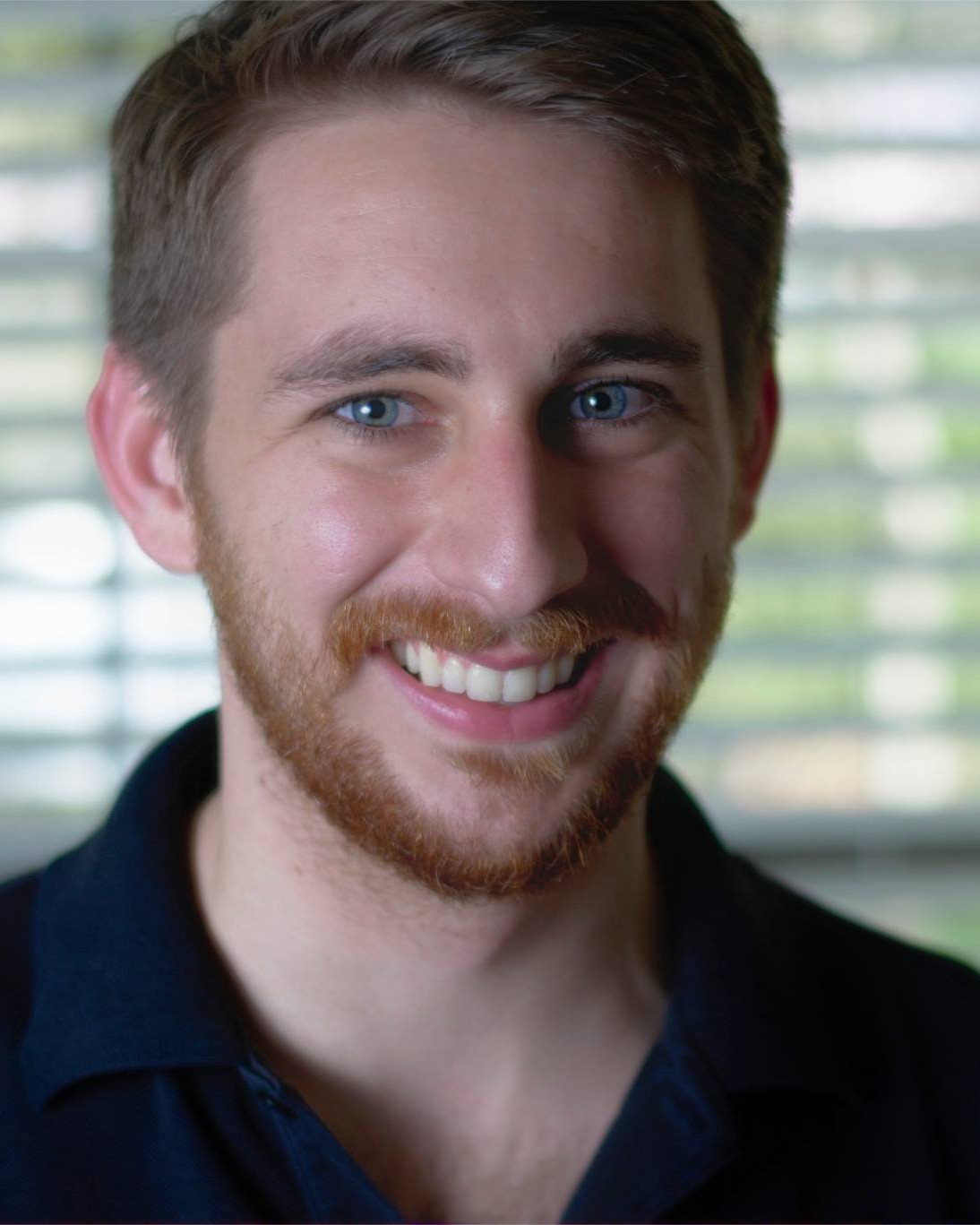}}]{Chance Tarver}
received a B.S. degree in electrical engineering from Louisiana Tech University, Ruston, LA, in 2014, and his M.S. degree in electrical and computer engineering in 2016 from Rice University, Houston, TX. He is currently a Ph.D. student in the Department of Electrical and Computer Engineering at Rice University. His research interests include software defined radio and signal processing for wireless communications.
\end{IEEEbiography}
\begin{IEEEbiography}[{\includegraphics[width=1in,height=1.25in,clip,keepaspectratio]{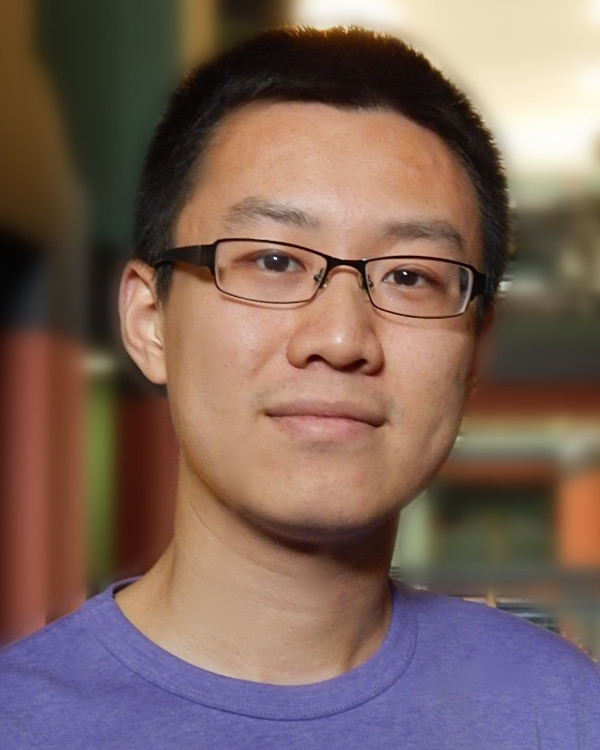}}]{Kaipeng Li}
(S'14) received his B.S. degree in physics from Nanjing University, Nanjing, China, in 2013, and his M.S. degree in electrical and computer engineering from Rice University in 2015. He is currently a Ph.D. candidate in the Department of Electrical and Computer Engineering at Rice University, Houston, Texas. His research interests include digital signal processing, parallel computing on GPGPU and multicore CPU, software-defined radios and massive MIMO systems.
\end{IEEEbiography}
\begin{IEEEbiography}[{\includegraphics[width=1in,height=1.25in,clip,keepaspectratio]{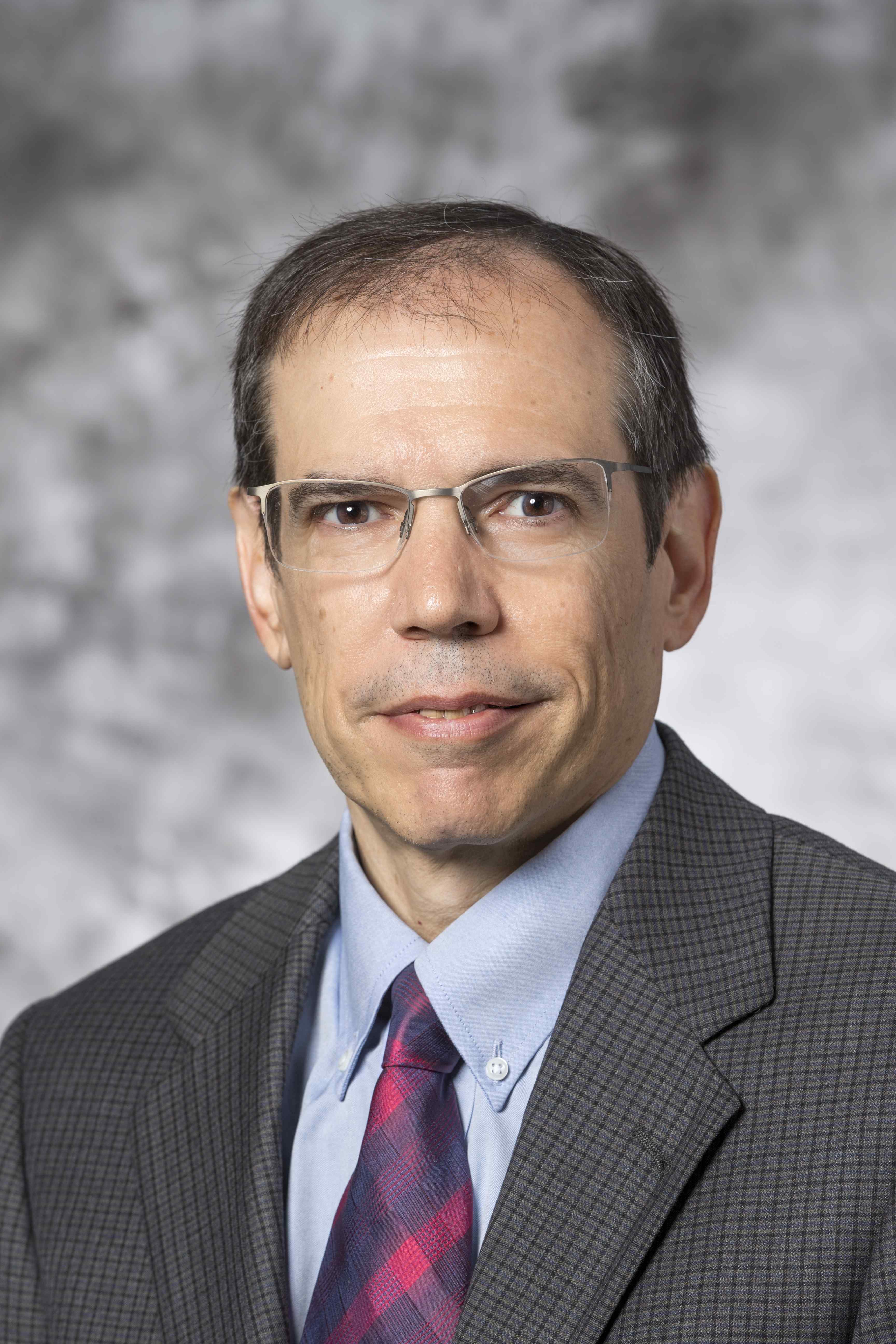}}]{Joseph R. Cavallaro}
(S'78, M'82, SM'05, F'15) received the B.S. degree from the University of Pennsylvania, Philadelphia, PA, in 1981, the M.S. degree from Princeton University, Princeton, NJ, in 1982, and the Ph.D. degree from Cornell University, Ithaca, NY, in 1988, all in electrical engineering. From 1981 to 1983, he was with AT$\&$T Bell Laboratories, Holmdel, NJ. In 1988, he joined the faculty of Rice University, Houston, TX, where he is currently a Professor of electrical and computer engineering. His research interests include computer arithmetic, and DSP, GPU, FPGA, and VLSI architectures for applications in wireless communications. During the 1996-1997 academic year, he served at the National Science Foundation as Director of the Prototyping Tools and Methodology Program. He was a Nokia Foundation Fellow and a Visiting Professor at the University of Oulu, Finland in 2005 and continues his affiliation there as an Adjunct Professor. He is currently the Director of the Center for Multimedia Communication at Rice University. He is a Fellow of the IEEE and a Member of the IEEE SPS TC on Design and Implementation of Signal Processing Systems and the Chair-Elect of the IEEE CAS TC on Circuits and Systems for Communications. He is currently an Associate Editor of the IEEE Transactions on Signal Processing, the IEEE Signal Processing Letters, and the Journal of Signal Processing Systems. He was Co-chair of the 2004 Signal Processing for Communications Symposium at the IEEE Global Communications Conference and General/Program Co-chair of the 2003, 2004, and 2011 IEEE International Conference on Application-Specific Systems, Architectures and Processors (ASAP), General/Program Co-chair for the 2012, 2014 ACM/IEEE GLSVLSI, Finance Chair for the 2013 IEEE GlobalSIP conference, and TPC Co-Chair of the 2016 IEEE SiPS workshop. He was a member of the IEEE CAS Society Board of Governors during 2014. 
\end{IEEEbiography}
\begin{IEEEbiography}[{\includegraphics[width=1in,height=1.25in,clip,keepaspectratio]{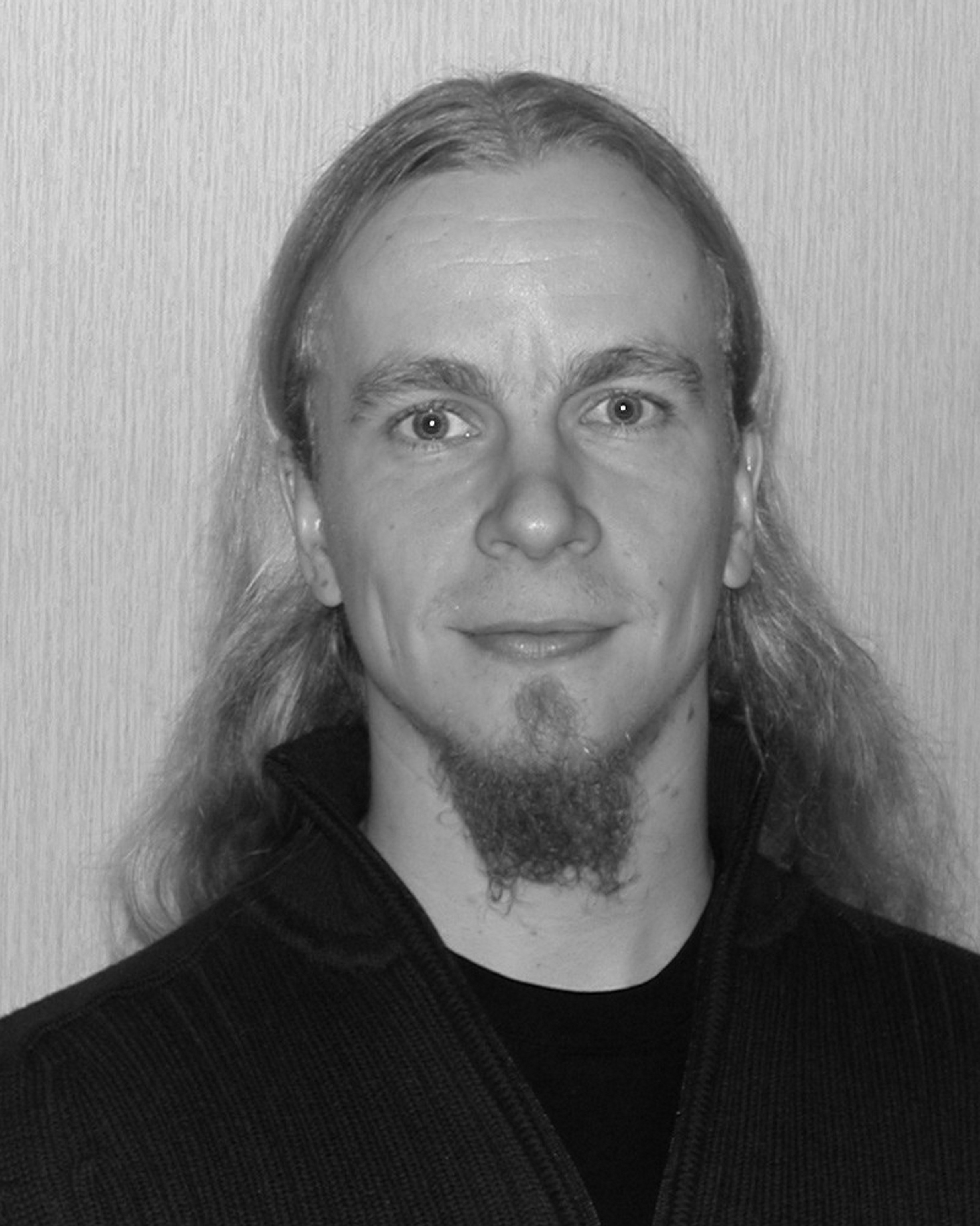}}]{Mikko Valkama}
(S'00, M'01, SM'15) was born in Pirkkala, Finland, on November 27, 1975. He received the M.Sc. and Ph.D. Degrees (both with honors) in electrical engineering (EE) from Tampere University of Technology (TUT), Finland, in 2000 and 2001, respectively. In 2002, he received the Best Ph.D. Thesis -award by the Finnish Academy of Science and Letters for his dissertation entitled "Advanced I/Q signal processing for wideband receivers: Models and algorithms". In 2003, he was working as a visiting researcher with the Communications Systems and Signal Processing Institute at SDSU, San Diego, CA. Currently, he is a Full Professor and Department Vice-Head at the Department of Electronics and Communications Engineering at TUT, Finland. His general research interests include communications signal processing, estimation and detection techniques, signal processing algorithms for software defined flexible radios, cognitive radio, full-duplex radio, radio localization, 5G mobile cellular radio, digital transmission techniques such as different variants of multicarrier modulation methods and OFDM, and radio resource management for ad-hoc and mobile networks.
\end{IEEEbiography}

\end{document}